\documentstyle[11pt]{article}
\begin{document}
\vspace{1cm}

\begin{center}
{\bf\Huge
Resonance trapping and saturation\\

\vspace{0.5cm}

of decay widths}

\vspace*{1cm}

{\bf \Large E.~Persson, T.~Gorin and I.~Rotter}\\

\vspace{0.5cm}

{\it
Max-Planck-Institut f\"ur Physik komplexer Systeme,\\
D-01187 Dresden, Germany, {\rm and}\\
Technische Universit\"at Dresden, Institut f\"ur Theoretische Physik,\\
D-01062 Dresden, Germany \\
}
\vspace{0.5cm}
{\small
 persson,gorin,rotter@mpipks-dresden.mpg.de}

\end{center}

\vspace*{2cm}

\begin{abstract}

Resonance trapping appears in open many-particle quantum systems at high 
level density when the coupling to the continuum of decay
channels reaches a critical strength. Here a reorganization of the
system takes place and
a separation of different time scales appears.
 We investigate it under the influence of additional
weakly coupled channels as well as by
taking into account the real part of the coupling term between system and
continuum. We observe a saturation of the mean width of the
trapped states. Also the decay rates saturate as a function of the
coupling strength. The mechanism of the saturation is studied in detail.
In any case, the critical region of reorganization is enlarged.
When the transmission coefficients for the different channels are
different, the width distribution is broadened as compared
to a $\chi_K^2$ distribution where $K$ is the number of channels.
Resonance trapping  takes place before the broad state overlaps
regions beyond the extension of the spectrum of the closed system.\\

\end{abstract}

\newpage

\section{Introduction}

At low excitation energy, the states of a many-particle quantum system
are usually well isolated from one another.
Their coupling  via the continuum of decay channels
can therefore be  neglected. The reaction cross
section consists of a sum of resonances with Breit-Wigner shapes the
positions, widths and partial widths of which are well defined.  \\

At higher excitation energy both the level density and the decay widths of
the states in most systems become so large  that the resonance states
overlap. The cross section
is an interference picture. Neither the positions, the widths nor the
partial widths of the resonance states can unambiguously be determined from
an analysis of  the cross section, see e.g. \cite{sorosamu}.
Meaningful values can be
obtained  from time measurements which provide the average lifetime of
the resonance states at high level density. Using the channeling
method, the mean lifetime of fine structure resonances under isobaric
analogue resonances in medium nuclei was found experimentally  to be much
longer than expected on the basis of the statistical
theory of nuclear reactions \cite{temmer}. The modification of the Fano
profiles of autoionizing states due to the coherent coupling with each
other has been studied experimentally by means of the two states $3p^2~^1S$
and $3p3d~^1P$ in magnesium atoms \cite{atom}  \\

Theoretically, the transition from low  to high level density is studied
recently in different papers for different systems, see \cite{sorosamu}
and \cite{klrousw} to \cite{gorin}.
 As a function of $\bar\Gamma / D$ (where
$\bar\Gamma $ is the average value of the widths of all resonance states,
being a measure for the  coupling strength of the system to the
continuum, and $D$ is the mean level distance) all results show the
same characteristic features: beyond a critical value of $\bar\Gamma / D$,
separated time scales exist when the number of channels
is smaller than the number of resonance states. The very existence of
different time scales at high level density corresponds to the basic
assumption of the unified theory of nuclear reactions formulated
phenomenologically about 40 years ago by Feshbach \cite{fesh} for many
resonances coupled to a small number of common decay channels.\\

The main differences in the results of the different theoretical approaches
consist in the behavior of the long-lived resonance states as a function of
further increased  $\bar\Gamma / D$. While the widths saturate in the more
realistic models for many-body systems
 e.g. \cite{ismuro}, they decrease in the random matrix
models. A saturation of the decay rates is observed
also in the ``bottle-neck'' picture of transition state theory
  which relates the saturation value to the number of
independent decay channels rather than to the widths of individual
resonance states of the system \cite{miller}. As a reply to \cite{comment},
the mechanism of the saturation is shown by the same authors to be
associated with a broadening of the distribution of resonance widths
\cite{reply}.  A broadening of the width distribution at high level
 density is found also, e.g., in \cite{miller2,taylor} and shown in
\cite{ro91,dirose,levrep,keff,fyod,gorin} to be
caused by resonance trapping
which is the basic process of the redistribution of the system. It leads
ultimately to the formation of different time scales.\\

In this paper, we study in detail the widths of the long-lived states
as a function of the coupling strength to the continuum since
this is a controversial point of discussion. Most
calculations are performed beyond the standard random matrix approach.\\

In sect.~2, the phenomenon of resonance trapping is described in
the framework of the random matrix theory. The widths of the trapped
resonance states do not saturate  with increasing coupling strength
to the continuum. In sect.~3, the    
 spectroscopic values of an open many-particle quantum 
system are given. Additonal terms in the effective Hamiltonian 
appear which may prevent the widths of the trapped resonance states to 
decrease with increasing coupling to the continuum. 
In the following sections, we discuss the influence 
of additonal terms in the effective Hamiltonian
on the widths of the long-lived trapped states.
 Additional weakly coupled channels  are shown to cause a saturation
of the mean width with further increasing coupling strength (sect.~4).
The sharp distinction between short- and long-lived
states is removed by the real part of the coupling term to the
continuum (sect.~5). In both cases, further avoided crossings
appear and  the bi-orthogonality of
the eigenfunctions of $H^{\rm eff}$ continues to be essential.
That means, the critical region where the redistribution takes
place is enlarged. This can be seen also in the
width distribution of the trapped states, investigated in sect.~6.
 Some conclusions on the
broadening of the width distribution, the separation of time scales and
the saturation of decay rates at high level density are drawn in the
last section.\\

\section{Random matrix theory and resonance trapping}

\subsection{Basic equations}
\label{sec:rmm}

In order to describe resonance phenomena at high level density in a
small energy interval far from thresholds the random matrix theory
has been developed. 
The effective Hamiltonian is (see e.g. \cite{mawei,soze,pave})
\begin{eqnarray}
{H_{\rm RMM}^{\rm eff}}=H_{\rm b}-iW=H_{\rm b}-\frac{i}{2} VV^T \; .
\label{eq:hrmm}
\end{eqnarray}
Here $V$ is an energy independent random matrix consisting of $K$ random
vectors of dimension $N$ with matrix elements $V_R^c$.
 The value
$V_c^2= \sum_{R=1}^N (V_R^c)^2$
gives the coupling strength of the system to the channel $c$.
Each vector $V_c$ has Gaussian
distributed elements with mean value $0$ and variance $V_c^2/N$.
$H_b$ is chosen from  the Gaussian Orthogonal Ensemble (GOE)
and the  mean level density follows a semi-circle law. In our case
the length  of the spectrum is $L=2$ units and therefore the mean level
distance in the middle of the spectrum is given by $d=\pi/(2N)$.\\

According to \cite{mol,sim}, one can obtain the mean width
$\langle\Gamma_l\rangle$ of the long-lived resonance states
by considering the diagonal elements of the $S$-matrix, averaged
over a sufficiently large energy interval. It holds
\begin{eqnarray}
\prod_{c=1}^K |\langle S_{cc}\rangle| =
{\rm e}^{-\pi \langle\Gamma_l\rangle/d} \; .
\label{eq:mosi}
\end{eqnarray}
The mean width $\langle\Gamma_l\rangle$ of the states in the middle
of the spectrum is
\begin{eqnarray}
\langle\Gamma_l\rangle = -\frac{d}{2\pi} \sum_{c=1}^{K} \ln (1-\tau_c) \; ,
\label{eq:mosi2}
\end{eqnarray}
where the $\tau_c$ are the
transmission coefficients, defined as follows,
\begin{eqnarray}
\tau_c = 1- |\langle S_{cc}\rangle|^2 \;.
\label{eq:taudef}
\end{eqnarray}
These coefficients are calculated in \cite{agassi} using a resummation
method in the power series expansion of the transfer matrix. The result is
\begin{eqnarray}
\tau_c = \frac{4 x_c}{(1+ x_c)^2} \; ,\qquad
x_c = \frac{\pi V_c^2}{2 Nd} \;.
\label{eq:mosi1}
\end{eqnarray}

The mean width $\bar\Gamma$  of {\it all} resonance states can be
obtained directly from (\ref{eq:hrmm}),
\begin{eqnarray}
\bar\Gamma = \frac{2}{N}
{\rm Im} \;
\{ {\rm tr} \;
H_{\rm RMM }^{\rm eff} \} =
\frac{1}{N}\sum_c V_c^2 \;
= \frac{2d}{\pi}\sum_c~x_c \; .
\label{eq:gtot}
\end{eqnarray}
The parameters $x_c$ measure the coupling strengths of the system to the
different decay
channels $c$. For increasing total coupling they subsequently pass over the
critical value one. Here the corresponding transmission coefficient
reaches its maximal value $\tau_c=1$.
The logarithm in (\ref{eq:mosi2}) diverges and
Eq.~(\ref{eq:mosi2}) looses its  validity.
Eq.~(\ref{eq:mosi1}) is symmetric with respect to an
exchange of $x_c$ by $1/x_c$.
As a consequence the transmission coefficients do
not distinguish between weak and strong total coupling strength.
We have $\tau_c<1$ for all $x_c\ne 1$.
\\

In the following sections we will see, that at each of the points
$\tau_c = 1$
a single  resonance state separates in width from the remaining ones.
For $x_c>1$ this  state becomes much broader than the other
 ones and can
no longer be described statistically together with the long-lived resonance
states.
The width of the short-lived  state is \cite{diharo,fyso}
\begin{eqnarray}
\Gamma_s = \frac{2 N d}{\pi} \left(x_c - \frac{1}{x_c}\right) = 
\left(x_c - \frac{1}{x_c}\right) \; ; \; \; \;  x_c > 1   
\label{eq:wshort}
\end{eqnarray}
while the widths of the long-lived states decrease as $1 / x_c$.
\\

\subsection{Resonance trapping at strong coupling}
\label{sec:trapp}

One interesting question is how the system described by the hamiltonian
$H_{\rm RMM}^{\rm eff}$ behaves as the
coupling to the continuum increases. To investigate this question,
we replace  $V$ by $\alpha V$
 in the term $W$ of the effective Hamiltonian, Eq. (\ref{eq:hrmm}),
and vary $\alpha$.
\\

For small $\alpha$ we can treat $W$  as a small perturbation on
 $H_b$. The rank of $H_b$ is N.
${H^{\rm eff}_{RMM}}$ thus describes $N$ states with energies determined
by the eigenvalues of $H_b$. The widths are proportional to
$\alpha ^2$. This holds well as long as the resonances are non-overlapping.\\

For large $\alpha$  we can treat  $H_b$ as a small
perturbation on $W$. As $W$ has rank $K$ it directly follows
that only $K$ states have large widths, see e.g. \cite{soze,pave}.
The widths of the remaining $N-K$ states are small. That means,
two different time-scales exist.\\

In this subsection, we first illustrate the phenomenon of resonance
trapping for $N$ resonance states coupled to $K=1$ open decay channel. 
We define
\begin{eqnarray}
\kappa^j=
\frac{2}{N-j}\sum_{R=j+1}^N\frac{\Gamma_R}{D_0}
=2~\frac{\bar\Gamma_{N-j}}{D_0} \; .
\label{equ:kappaj}
\end{eqnarray}
Here the sum runs over
all but the j broadest states. $D_0 = L/(N-1)$ is the mean level spacing
of the eigenvalues of $H_b$ and $L$ ist the length of the spectrum.
For $N-j \gg 1$ the difference
between $\kappa^{j-1}$ and $\kappa^{j}$ is $2\Gamma_j/(D_0(N-j))$
(states ordered according to decreasing width). Further we define
\begin{eqnarray}
\kappa^{\rm tot}=\kappa^{j=0}=\frac{2}{N}\sum_{R=1}^N
\frac{\Gamma_R}{D_0} =2 \frac{\bar\Gamma}{D_0} \; .
\label{eq:kappatot}
\end{eqnarray}
$\kappa^{\rm tot}$ is a measure of the total coupling strength of the system
to the continuum (see Eq. (\ref{eq:gtot})). In the 
random matrix theory (with $N$ large) we
have $d=\pi/(2N)$ and $L=2$ and thus $D_0=4d/\pi$ which gives
$\kappa^{\rm tot}=x_c$ in the one-channel case
(compare subsection \ref{sec:rmm}).\\

Within the random matrix theory, we calculate $\kappa^j$ versus
$\kappa^{\rm tot}$ for $N=300$, $K=1$ and $j=0,1,2$ with varying coupling
to the continuum (fig. 1.a). The curves shown are averages over $20$
calculations. Note the logarithmic scales.
Until $\kappa^{\rm tot} \approx 1$ the average width of all states
increases  with increasing coupling to the continuum.
At $\kappa^{\rm tot} \approx 1$ two globally
separated time-scales are formed. The broad state  should be identified
with a doorway state \cite{doorway} and a tight transition state
\cite{transition}, respectively.  At the separation point $\kappa^{\rm
tot}=1$, the transmission coefficient is $\tau_c=1$. For still further
increasing $\kappa^{\rm tot}$  the broadest state is getting a still larger
width but the average width of the remaining ones decreases.\\

The thick line in fig. 1.a. shows $2 \langle\Gamma_l\rangle / D_0$
obtained from Eqs. (\ref{eq:mosi2}) and (\ref{eq:mosi1}).  For $\kappa^{\rm
tot} < 1$, $\langle\Gamma_l\rangle$ is  
 the mean width 
of all $N$ resonance states, $\langle\Gamma_l\rangle \approx \bar\Gamma$, 
while for $\kappa^{\rm tot} >  1$,  
 $\langle\Gamma_l\rangle$ is  the mean width
of the $N-1$ trapped states, $\langle\Gamma_l\rangle
\approx \bar\Gamma_{N-1}$.
Therefore, the transmission coefficients $\tau_c$ related to the
$\langle\Gamma_l\rangle$ by Eq.  (\ref{eq:mosi2}) do not give us
information on the coupling strength $\kappa^{\rm tot}$ of the system to
the continuum (compare subsection \ref{sec:rmm}).\\

Resonance trapping of many states takes place whenever the local
level density compared to the local mean width is sufficiently large,
see e.g. \cite{ismuro,levrep,keff}.\\

To illustrate the local properties of resonance trapping we
show in fig. 1.b the widths $\Gamma_R$ versus the energies $E_R$ for the
states of fig. 1.a. Only a small part of the spectrum is shown.
The calculations are performed for $0.1\le\kappa^{\rm tot}\le 10$ in
steps of $\log \kappa^{\rm tot} =0.04$.
The points for $\kappa^{\rm tot}=0.1$, $1$ and $10$
are marked with triangles, stars and squares, respectively.
As a function of $\alpha$ the complex eigenvalue
$\varepsilon_R=E_R-\frac{1}{2}\Gamma_R$ of each
resonance state follows a certain "trajectory". For small 
coupling to the continuum,
the widths of the states increase with increasing continuum coupling.
This process takes place for every resonance state
 up to that value of the continuum coupling at which
 the state starts to overlap one of the resonance
states in its neighborhood. The crossing of resonance states is avoided
in the complex plane: The states attract each
other in energy and their widths bifurcate,
 i.e. the width of one of the states continues
to grow with further increasing coupling strength while that of the
 other one decreases.
The state finally being the broadest one goes through a number of
``collisions'' (avoided resonance crossing between two states looks like
a ``collision'' in this representation) before it dominates the complete
spectrum. It is formed in the middle of the spectrum
where the level density is the largest.\\

The stars on the trajectories (fig. 1.b) show that trapping of resonance
states
is a local process which takes place {\it before} different timescales
are formed globally. The widths of all the trapped states decrease with 
further increasing coupling strength. \\

This behaviour does not change when the level density  has a band or 
shell structure.
We simulate such a situation by choosing a $H_{\rm b}$,
Eq. ($\ref{eq:hrmm}$), with
several more or less separated regions of high level density. Between
them and at the borders of the spectrum, the level density is smaller.
In the present calculation we have $K=4$ and $N=300$.
The coupling of each state to each channel is randomly chosen, i.e. the
coupling strengths of the four channels are equal to one another.
We have three regions with Gaussian shaped
level density. Two of the regions are close to each other and the third
one is lying well separated from the other two.
The energies $E_R$ and widths $\Gamma_R$ of the resonance
states are shown in
fig. 2 for varying total coupling strength to the continuum.\\

We see the following result. In each group of states  there are formed
some broad states at high level density. These states do {\it not} overlap
regions outside the interval studied.
With further increasing coupling strength, these broad
states attract one another in energy and their widths bifurcate.
Finally, there are only four broad states according to the four
channels. In the separation point,
their widths are smaller than
the energy region covered by the long-lived states.
The question whether there are states lying outside the extension
of the spectrum is  of interest only when studying the behaviour of the
short-lived states at a further increased strength of the coupling to the
continuum.  The resonance trapping in every group  is left unchanged by the
effects at the borders of the spectrum.\\

We conclude this section by stating the following.  
In the random matrix theory, 
the widths of the trapped  resonance states {\it decrease}
  as a function of increasing coupling strength
$\kappa^{{\rm tot}}$. \\

\section{Continuum shell model (CSM) for an open many-body quantum system}
\label{sec:csm}

In many-body quantum
 systems, the widths of the long-lived resonance states do not
decrease but {\it saturate} as a function of the coupling strength 
$\kappa^{tot}$ \cite{comment,reply}. In order to investigate the mechanism of
saturation we have, according to the results shown in 
subsection  \ref{sec:trapp}, to go
beyond the random matrix theory. In the following,
we sketch a model which allows
 to describe open {\it many-body} quantum systems.\\

The time independent Schr\"odinger equation
\begin{eqnarray}
(H-E) \mid \Psi \rangle = 0
\label{eq:scr}
\end{eqnarray}
is solved
in a Hilbert space consisting not only of the discrete many-particle states
of a closed system but also of the continuum of decay
channels. The potential is assumed to be a spherical one.
The Hamilton operator of the  system is
\begin{eqnarray}
{H = H^0 + {\hat V} \; .}
\label{eq:totham}
\end{eqnarray}
$H^0$ is the unperturbed Hamilton operator describing particles in a finite
depth potential and ${\hat V}$ is the operator of the two-body residual
interaction between the particles.
For details see \cite{ro91}.
The relation between $V$, Eq.~(\ref{eq:hrmm}),
 and the two-body operator $\hat V$ is considered
in \cite{brody}.
\\

In order to find the solution $|\Psi\rangle$ of ($\ref{eq:scr}$) we
first solve the shell model problem
\begin{eqnarray}
(E_R^{sm}-H_{\rm QQ}) \mid \phi_R^{sm} \rangle = 0
\label{eq:smev}
\end{eqnarray}
in the $Q$ subspace of $N$ discrete states and the coupled channel equations
\begin{eqnarray}
{(E^{(+)}-H_{\rm PP}) \mid \xi_E^{c(+)} \rangle = 0}
\label{eq:copch}
\end{eqnarray}
with the proper boundary conditions in the $P$ subspace of
$K$ coupled channels. The projection operators are
${\hat Q = \sum_{R=1}^N \mid \phi_R^{sm}
\rangle \langle \phi_R^{sm} \mid}$
and
${\hat P = \sum_{c=1}^K \int dE \mid \xi_E^c \rangle
\langle \xi_E^c \mid .}$
 Here $H_{\rm QQ}\equiv \hat QH \hat Q$ and $H_{\rm PP}\equiv\hat PH\hat P$.
Further, we solve the coupled channel equations with source term
\begin{eqnarray}
(E^{(+)}-H_{\rm PP})|\omega_R^{(+)}\rangle=
{\hat V}_{\rm PQ}| \phi_R^{sm}\rangle \; ,
\label{eq:source}
\end{eqnarray}
which connects the two subspaces. Using
$\hat P+\hat Q=1$
we then express $|\Psi\rangle$ by means of the solutions of ($\ref{eq:smev}$),
($\ref{eq:copch}$) and ($\ref{eq:source}$). Care must be taken in order to
avoid double-counting 
from ($\ref{eq:smev}$) and
($\ref{eq:copch}$), i.e. appearance of any resonances in ($\ref{eq:copch}$).
For this purpose a cut-off technique for single particle
resonances is used in \cite{ro91} when solving ($\ref{eq:copch}$).\\

In $P$ subspace, the propagator is
$G_{\rm P}^{(+)}=\hat P (E+i\epsilon-H_{\rm PP})^{-1}\hat P \; .$
The propagator $\hat Q(E-{H_{\rm QQ}^{\rm eff}})^{-1}\hat Q$ in $Q$ subspace
contains the effective Hamilton operator in this subspace,
\begin{eqnarray}
{{H_{\rm QQ}^{\rm eff}}(E)=H_{\rm QQ}+{\hat V}_{\rm QP}G_{\rm P}^{(+)}
{\hat V}_{\rm PQ} \; .}
\label{eq:heff}
\end{eqnarray}
${H_{\rm QQ}^{\rm eff}}(E)$ is non-hermitean and energy dependent.
It has energy dependent complex eigenvalues
$\tilde{\varepsilon}_r=\tilde E_R-\frac{i}{2}\tilde\Gamma_R$
and eigenfunctions $\mid \tilde{\Phi}_R\rangle$ describing the quasi-bound
states embedded in the continuum (QBSEC) \cite{ro91}.
The $\mid \tilde{\Phi}_R\rangle$ form a bi-orthogonal set at each energy $E$
with the orthogonality relation 
$\langle {{\tilde \Phi}_{R}}^* | {{\tilde \Phi}_R'} \rangle
=  \delta_{RR'}\;$ (see \cite{keff}).
Further, it holds
$\langle {{\tilde \Phi}_{R}} | {{\tilde \Phi}_R} \rangle \geq 1\;,
 \; \langle {{\tilde \Phi}_{R}} | {{\tilde \Phi}_R'} \rangle \in C
 \; (R\not= R')\; .$
\\

Diagonalizing ${H_{\rm QQ}^{\rm eff}}$ we get the solution of
 ($\ref{eq:scr}$) as
\begin{eqnarray}
{\mid \Psi_E^{(+)} \rangle =  \mid \xi_E^{c(+)} \rangle +
\sum_{R=1}^N \mid \tilde{\Omega}_R^{(+)} \rangle
\frac{1}{E-\tilde{\varepsilon}_R}\langle \tilde{\Phi}_R \mid {\hat V}
 \mid \xi_E^{c(+)} \rangle \;.}
\label{eq:totsol}
\end{eqnarray}
Here the
\begin{eqnarray}
|\tilde\Omega^{(+)}_R\rangle=
(1+G_{\rm P}^{(+)}{\hat V})\mid \tilde{\Phi}_R\rangle
\label{eq:Omega}
\end{eqnarray}
are the wavefunctions of the resonance states R.
It should be stressed here once more that ${\hat V}$ is the operator of
the {\it two-body} residual interaction
and that the $\tilde \Phi_R$ are {\it many-particle} wavefunctions.\\

The relation ($\ref{eq:Omega}$) between the wavefunctions
$\tilde\Omega^{(+)}_R$ of the resonance states and the eigenfunctions
$\tilde\Phi_R$ of ${H_{\rm QQ}^{\rm eff}}$ is analogous to the
Lippman-Schwinger equation
\begin{eqnarray}
|\xi_E^c\rangle =(1+G_{\rm P}^{(+)}{\hat V})\; |\chi_E^c\rangle
 \label{eq:lippschw}
\end{eqnarray}
between channel wavefunctions $\chi_E^c$ and coupled channel
wavefunctions $\xi_E^c$. Therefore 
$\langle {\tilde \Omega}_R
  |{\hat V} | \chi_E^{c(+)} \rangle = 
\langle {\tilde \Phi}_R
|{\hat V} | \xi_E^{c(+)} \rangle\; .$
We define the amplitude of the partial width by
\begin{eqnarray}
 \tilde\gamma_{Rc}  \equiv 
\frac{1}{\sqrt{2\pi}}
\langle\tilde\Omega_R|{\hat V}|\chi^{c(+)}_{E}\rangle = 
\frac{1}{\sqrt{2\pi}}
\langle{{\tilde \Phi}_R}|{\hat V}|\xi^{c(+)}_{E}\rangle \; .
\label{eq:apwidth}
\end{eqnarray}
By inserting the expression ($\ref{eq:totsol}$) for $\Psi$ into
the S-matrix
and using ($\ref{eq:apwidth}$) we get
\begin{eqnarray}
S_{cc'}=
e^{2i\delta_c} \delta_{cc'} -
2i\pi \langle \chi^{c'(-)}_{E}|{\hat V}|\xi_E^{c(+)}\rangle +
 i \sum_R  \frac{\tilde \gamma_{Rc} {\tilde \gamma}_{Rc'}}
{E - {{\tilde E}_R} + \frac{i}{2} {{\tilde \Gamma}_R}} \;.
\label{eq:smat}
\end{eqnarray}
The first two terms in ($\ref{eq:smat}$) describe the direct
reaction part of the process. The last term describes the resonance part,
i.e. excitation and de-excitation of the resonance states R.
The resonance part $ S_{cc'}^{res}$ of the $S-$matrix has the standard
Breit-Wigner form.
Note however that $\tilde\gamma_{Rc}$ is complex and energy dependent.
Also $\tilde E_R$ and $\tilde\Gamma_R$ are energy dependent functions.
$S_{cc'}^{res}(E)$
contains the contributions of all the resonance states at the energy
$E$ of the system. These contributions may be very different from
those at other energies (see \cite{thres}).\\

The poles of the S-matrix give the energies $E_R$
and widths $\Gamma_R$
of the resonance states. They are defined by the fixed-point equations
$E_R = \tilde E_R(E=E_R)$ and $\Gamma_R = \tilde \Gamma_R(E=E_R)$.
The number of resonance states is exactly equal to the number $N$ of discrete
states obtained from ($\ref{eq:smev}$) if
double counting in $P$ and $Q$ is avoided.\\

If the energy dependence of ${{\tilde E}_R}(E)$ and
${{\tilde \Gamma}_R}(E)$ is weak in the interval studied, it holds
$E_R\approx{{\tilde E}_R}(E')$ and $\Gamma_R\approx{{\tilde \Gamma}_R}(E')$
where $E'$ is an energy somewhere in the middle of the interval.
Only under such conditions, $E_R$ and $\Gamma_R$ can be considered as
resonance parameters. Otherwise, the spectroscopic studies performed at the
energy $E'$ are meaningful only at this energy since only here
the orthogonality relations between the right and left
wavefunctions are fulfilled.
The present calculations are performed at the energy $E = E'$ of the system.\\

According to \cite{keff}, the decay rates can be defined by
$k^{\rm eff} (t) = -\frac{d} {d t} \;
\ln  \langle \phi(t) | \phi(t) \rangle $
where $\phi(t)$ is the wave function  of the system.
Using the ansatz
$| {\phi} (t) \rangle = \sum_R a_R(t) \; | {{\tilde \Phi}_R} \rangle$
and solving the time-dependent Schr\"odinger equation
with the effective Hamiltonian $H_{\rm QQ}^{\rm eff}$ we get 
\begin{eqnarray}
| \phi(t) \rangle =
\sum_R a_R(0) e^{- \frac{i} {\hbar}
(\tilde E_R - \frac{i} {2} \tilde \Gamma_R) \; t} \; |
{{\tilde \Phi}_R} \rangle \; .
\label{eq:high1}
\end{eqnarray}
The $a_R(0)$ define the wavefunction $\phi(0)$ of the system at
the time $t=0$.
Neglecting the oscillations caused by the bi-orthogonality
of the function system
we get
\begin{eqnarray}
k^{\rm eff}_{\rm gr}(t) & = &
\frac{1}{\hbar} \; \frac{\sum_R A_R^2(t) \; \tilde\Gamma_R}
{\sum_R A_R^2(t)}
\; \equiv \; \frac{1}{\hbar} \bar\Gamma^{(A)}
\label{eq:rate6}
\end{eqnarray}
in analogy to $k^{\rm eff}_R=\tilde\Gamma_R/\hbar$ for isolated
resonances.
The $A_R^2(t)$ are given by
$A_R^2(t)=|a_R(0)|^2 $ $ e^{-\tilde \Gamma_R \; t / \hbar }\;
\langle\tilde \Phi_R |\tilde \Phi_R\rangle  \; .$
They decrease exponentially with the rate
$\tilde\Gamma_R/\hbar$. Therefore the $A_R^2(t)$ for the
short-lived states are neglible in the long-time scale and the
sum in (\ref{eq:rate6}) runs only over the trapped states.
If the width distribution of the $N-K$ trapped states is narrow,
it holds $\bar\Gamma^{(A)}\approx\bar\Gamma_{N-K}$.
The weighted width $ \bar\Gamma^{(A)}$ is
generally time dependent.\\

Let us write Eq. ($\ref{eq:heff}$) more explicitly,
\begin{eqnarray}
{H_{\rm QQ}^{\rm eff}}(E)=H_{\rm QQ}+{\cal P}(E)-i W(E) \; .
\label{eq:heff2}
\end{eqnarray}
Here
\begin{eqnarray}
W_{RR'}(E)= \pi \sum_c  V_R^c (E) V_{R'}^c (E)
\label{eq:wdef}
\end{eqnarray}
where
$V_R^c(E)=\langle\phi_R^{sm}|{\hat V}|\chi^{c(+)}_{E}\rangle$ are
real numbers describing the coupling of the  shell model
states R to the channels c at the energy E of the system and $c$
runs over all open channels. Further,
\begin{eqnarray}
{\cal P}_{RR'}(E)=\sum_c P \int\limits_{\epsilon_c}^{\infty}
V_{R}^c (E') \frac{1} {E-E'} V_{R'}^c (E')dE' \; .
\label{eq:pvdef}
\end{eqnarray}
$P$ denotes the principal value of the integral and $\epsilon_c$ is
the threshold energy for the channel c. As a rule, the 
$\epsilon_c$ are different for different channels.
\\

Characteristic of the Hamiltonian of an open many-body quantum system is
therefore (i) the different channels are coupled with different strength and
(ii) the coupling via the continuum contains not only an imaginary part but
also a real part. In the following sections we study the widths of trapped
resonance states as a function of increasing coupling strength $\kappa^{{\rm
tot}}$ by considering these properties of the Hamiltonian.\\

\section{Influence of additional weakly coupled channels}
\label{sec:dnc}

In a many-body quantum system, each channel has
a certain coupling strength which may be quite different from the coupling
strength of other channels. Sources for the different coupling strengths of the
channels are, above all, the structure of the states of the residual system,
the different angular momenta and the different threshold energies
$\epsilon_c$, see sect. 3.\\

 We  study the influence of different coupling
strengths of the channels onto the resonance trapping
in a schematical manner. The main emphasis
lies on the question whether the mean width of the trapped states at
strong coupling to the continuum saturates or approaches zero.
The study is performed in the
random matrix model (RMM) with the Hamiltonian 
($\ref{eq:hrmm}$)
and $N=300$, $K=4$,
 but different average coupling strength $V_c^2$
to the channels. 
The ratios among the coupling strengths are
${\rm V_{\rm c}^2}/{\rm V_{\rm c=1}^2}=1$, $0.1$, $0.01$ and $0.001$
(fig. 3.a) and
${\rm V_{\rm c}^2}/{\rm V_{\rm c=1}^2}=1$, $0.01$, $0.0032$ and $0.001$
(fig. 3.b). We show
$\kappa^j$ for $j = 0 ... 5$ averaged over $20$ calculations.\\

The separation points are defined  by $\tau_c=1$ for any channel,
i.e. $x_c=1$ for a certain $c$ (see Eqs. (\ref{eq:mosi2})
and (\ref{eq:mosi1})). These points are given by
$\kappa^{\rm tot}= \sum_{c'=1}^K {\rm V_{\rm c'}^2}/{\rm V_{\rm c}^2}$.
This gives $\kappa^{\rm tot}=1.11$, $11.1$, $111$, $1110$ in fig.
3.a and $\kappa^{\rm tot}=1.01$, $101$, $321$, $1010$ in fig. 3.b.\\

In fig.~3.a, there is at $\kappa^{\rm tot} \approx 1.2 $
a separation point, where one broad state separates from the other
ones and the mean width of the remaining resonance states starts to decrease.
Here, $\kappa^j$ for $j>0$  decreases slightly, but soon increases  again
under the influence of the next channel. At $\kappa^{\rm tot} \approx 12$ a
second broad state separates from the other ones. Similar situations
occur at  $\kappa^{\rm tot} \approx 120$ and
$\kappa^{\rm tot}\approx 1200$. Finally
there are four broad states corresponding to the four channels.
For still larger values of  $\kappa^{\rm tot}$,
the values $\kappa^j$, $j>3$ decrease.
Note that the values $\kappa^j$, $j>4$
are almost constant in the  region $1 < \kappa^{\rm tot} <
1500$. Thus, different coupling strengths to the channels
are a source for saturation of the average width of the
long-lived $N-K$ states as a function of $\kappa^{\rm tot}$.\\

In fig. 3.b, we have one strongly coupled channel and a group of weakly
coupled ones. At $\kappa^{\rm tot}\approx 1.1$
a picture similar to ordinary trapping with one channel
can be seen. One state separates and the widths of the remaining
ones decrease. For larger coupling to the continuum however,
the widths of the trapped states start to increase under the influence
of the weakly coupled channels. A new critical region occurs
between $\kappa^{\rm tot}\approx 70$ and $2000$ where
three states corresponding to the three new channels separate.
In this region, the widths of the trapped states saturate and thereafter
their widths  decrease.\\

Further, these two examples show another interesting result. At the points
where a broad state separates from the remaining ones, we have
$\tau_c\approx 1$ for a certain
channel c, see the thick line ($2 \langle\Gamma_l\rangle / D_0$
versus $\kappa^{\rm tot}$).
When all $\tau_c<1$, we have good agreement between
the $\langle\Gamma_l\rangle$ calculated
from Eqs. (\ref{eq:mosi2}) and (\ref{eq:taudef}) and the calculated
mean width $\bar\Gamma_{N-j}$. Here, $j$ is determined by the number of broad
states and $j=0$ only for $\kappa^{\rm tot}<1$ (compare subsection
\ref{sec:trapp}).\\

A small part of the eigenvalue picture ($E_R$ and $\Gamma_R$), fig. 3.c,
(${\rm V_{\rm c}^2}/{\rm V_{\rm c=1}^2}=1$, $0.01$,
$0.0032$ and $0.001$ as in fig.~3.b, 
but only one calculation) shows width increase, energy shift, width decrease
and once again increase, shift and decrease. The renewed increase of
the widths of the states trapped by the first channel is  caused by the fact
that the new channels become active only at strong coupling to the continuum.\\

 In fig.~3.d, we show
$\Gamma_R / 2$ and  $(x_c - 1/x_c) / 2 \; , \; c = 1, ... , 4 \; $,
Eq.~($\ref{eq:wshort}$),  versus 
$\kappa^{\rm tot}$ for the same calculation as in fig.~3.c.
The agreement between $(x_c - 1/x_c)$  and the four 
different widths $\Gamma_s$ of the short-lived states
 illustrates nicely that the separation of
every broad mode takes place from the group of long-lived resonance states.
This means, the separation process is
more or less independent of the
 broad modes separated at smaller values of $\kappa^{\rm tot}$.\\

\section{The role of the real part of the coupling term}
\label{sec:pvinf}

\subsection{Two resonances coupled to one common channel}
\label{sec:twores}

The influence of a real part  
onto the phenomenon of resonance trapping  
can be seen  by means of the following simple
model for two resonance states coupled to one common decay channel,
\begin{eqnarray}
h^{\rm eff} =
 \left(
\begin{array}{cc}
 1 &  0 \\
 0  & -1
\end{array}
\right)
- 2i \alpha e^{-i\theta}
\left(
\begin{array}{cc}
\cos ^2 \varphi & \cos \varphi \sin \varphi \\
\cos \varphi \sin \varphi & \sin ^2 \varphi
\end{array}
\right);
\;\;\; -90^0\le\theta\le 90^0    \; .
\label{eq:twol1}
\end{eqnarray}
Here the relative coupling strength of the two states to the continuum
may be varied by means of the angle $\varphi$
$\;(\varphi\neq 0^0, 90^0, ...)$. The angle $\theta$ determines
the ratio between the real and imaginary part of the coupling term.
In \cite{levrep} the case $\theta=0$ has been studied.\\

The eigenvalues of $h^{\rm eff}$ are
\begin{eqnarray}
\varepsilon _{\pm} = - i \alpha e^{-i\theta}
\pm \sqrt{1 - 2 i \alpha e^{-i\theta} \cos(2 \varphi) -
\alpha^2  e^{-2i\theta}} \; .
\label{eq:twol2}
\end{eqnarray}
At $\alpha = 0$ we have two states lying at the energies $-1$ and $1$. Their
widths increase with increasing $\alpha$ up to some critical value
$\alpha_{\rm crit} \le 1$.
The eigenvectors of $h^{\rm eff}$ are
\begin{eqnarray}
\Phi_\pm =
\frac{1}{\sqrt{-\alpha^2~e^{-2i\theta}~\sin^2(2\varphi)
+\phi_\pm^2}} \; \;
\left(
\begin{array}{c}
i\alpha~e^{-i\theta}~\sin(2\varphi)\\
\phi_\pm\\
\end{array}
\right) \; .
\label{eq:twol3}
\end{eqnarray}
Here, the normalization of the wavefunctions is made according
to $\langle{\Phi_\pm}^* | \Phi_\pm\rangle=1$
and
$\phi_\pm=
1-2i\alpha~e^{-i\theta}~\cos^2(\varphi)-\varepsilon_\pm\; .$
In our case of only two resonance states, the values $|\Phi_\pm|^2$ are 
the same for both states.\\

The distance in the complex plane between the two eigenvalues is
\begin{eqnarray}
|\varepsilon_+ - \varepsilon_-| =
2\left| \sqrt{1 - 2 i \alpha e^{-i\theta} \cos(2 \varphi) -
\alpha^2  e^{-2i\theta}}\right| \equiv 2 S
\label{eq:twodist}
\end{eqnarray}
Using $S$, the denominator in Eq. (\ref{eq:twol3}) can be rewritten
as follows,
\begin{eqnarray}
\sqrt{-\alpha^2~e^{-2i\theta}~\sin^2(2\varphi)+\phi_\pm^2}=
\sqrt{2S\left(S~\pm(1-ie^{-i\theta}\cos(2\varphi))\right)} \; .
\label{eq:tworseq}
\end{eqnarray}
The factor $(S~\pm(1-ie^{-i\theta}\cos(2\varphi)))$ is never zero
for finite $S$, and
thus  $|\Phi|^2\to\infty$
if and only if the distance in the complex plane between the two
eigenvalues goes to zero.\\

Let us define $\kappa=(\Gamma_1+\Gamma_2)/L$ where $L=2$ is the
distance between the two resonance states. For the hamiltonian
($\ref{eq:twol1}$) it holds $\kappa=2~\alpha~ \cos\theta$.\\

In the following the two cases $\varphi=22.5^0$ and $\varphi=45^0$
are studied in detail. In the first case both states coupled with equal
strength to the decay channel while
in the second case the state having the initial energy $+1$ is stronger
coupled to the channel than the other one.\\

In fig. 4 the complex eigenvalues (4.a) and
the $|\Phi|^2$ versus $\kappa$  (4.c) are shown for $\varphi=45^0$ and
$\theta=0^0$, $10^0$, $30^0$ and $45^0$. The different points in
fig. 4.a corespond to different values of $\kappa$, $0.2\le\kappa\le 20$.
As $\kappa$ grows, each eigenvalue follows a certain trajectory.
For $\theta=0^0$ the $E_R$ and
$\Gamma_R$ of the two states meet in one  point in the complex plane
at $\kappa=\kappa_{\rm crit}=2$. Here
$|\Phi(\kappa_{\rm crit})|^2 \to \infty$. Beyond this separation
point, one state continues to increase in width whereas the width of the
other one decreases ({\it resonance trapping}).
For $\theta > 0$ the state with initial energy $-1$ becomes the broader one
and is shifted towards negative energies.
In contrast to this,
for all $\theta<0$ (not shown in the figures) the state with inital energy $+1$
becomes, in this symmetrical case $\varphi=45^0$,
the broader one and is shifted towards large positive energies.
For $\theta\neq 0^0$ the minimum distance in the complex plane between
the two states remains different from zero and
$|\Phi|^2$ remains finite for all $\kappa$ with its maximum value at
$\kappa_{\rm crit}$. As $\theta\to 90^0$, $h^{\rm eff}$ becomes hermitian.
In this limit, $|\Phi|^2\to 1$ and $\kappa_{\rm crit}\to 0$.\\

The case $\varphi=22.5^0$ is shown in figs. 4.b (complex eigenvalues)
and 4.d ($|\Phi|^2$ versus $\kappa$) for some values of $\theta$
ranging from $-45^0$ to $+60^0$. For $\theta>0$ the state with initial energy
$E_R=+1$, being the broader one at small $\kappa$, gets
an extra shift towards small energies whereas the shift is towards large
energies for $\theta < 0$. $\kappa_{\rm crit}$ is a function of only
$|\theta|$ and has the same values as
in the symmetrical case. It  has its largest value for $\theta=0^0$.
At $\theta=45^0$ the complex eigenvalues of the
two states meet in one point and $|\Phi(\kappa_{\rm crit})|^2 $ diverges. As
$\theta\to\pm 90^0 ~ h^{\rm eff}$ becomes hermitean
and $|\Phi(\kappa^{\rm crit})|^2 \to 1$.
For $-90^0<\theta<45^0$ the state with initial energy $E_R=-1$
is the one becoming trapped. For $45^0<\theta<90^0$, however, the state with
initial energy $E_R=+1$ becomes trapped even
though that state is the broader one at small $\kappa$.\\

These examples show that the details of the resonance trapping change
when allowing for extra energy shifts by introducing
the angle $\theta$ in ($\ref{eq:twol1}$). They are basic for an understanding
of the results of the following section.
\\

\subsection{N resonance states coupled to K common channels}

We study the influence of the real part
of the coupling term onto the mean width of the trapped states
in the framework of the CSM. We do this by comparing the results of
calculations with and without ${\cal P}$,
Eq.~($\ref{eq:heff2}$), taken into account.\\

The rank of ${\cal{P}}$ is, generally, larger than $K$.
${\cal{P}}$ increases with $\alpha$ by approximatively the same
factor as $W$. Thus the hermitean part
$H_{\rm QQ}+{\cal{P}}$ of ${H_{\rm QQ}^{\rm eff}}$ is never
a small perturbation on $W$ and the rank of $H_{\rm QQ}^{\rm eff}$
is larger than $K$ also in the strong coupling
limit (compare sect.~\ref{sec:trapp}). Nevertheless, resonance
trapping occurs also when taking ${\cal{P}}$ into account.\\

In fig. 5 we present $\kappa^j$ versus $\kappa^{\rm tot}$
for some calculations in the  CSM.
As in the RMM, we use $D_0=L/(N-1)$ where $L$ is the length of the spectrum
of $H_{\rm QQ}$.
We study 190 resonance states in $^{16}O$ with $J^\pi = 1^-$ and vary $\alpha$.
In the first case (figs. 5.a,~b) we study the reaction
$^{15}O_{1/2^-}(n,n)^{15}O_{1/2^-}$ giving $K=2$ channels with
$s$ and $d$ waves, respectively. The second case (figs. 5.c,~d) is
$^{15}N_{1/2^-}(p,x)Y$ where $x=p,~n$ and
$Y=$ $^{15}N_{1/2^-,~3/2^-}$, $^{15}O_{1/2^-,~3/2^-}$
which gives  $K=10$ channels.
Figs. 5.a,~c and 5.b,~d are without and with ${\cal P}$, respectively.\\

The figures 5.a,~c with ${\cal P} = 0$  are similar to figs. 3.a,~b.
 In  fig. 5.a
we have 2 channels with different coupling strength. First we see the
increase of the mean   width of all the states for small coupling
 to the continuum
up to the separation point corresponding to the strong channel. Thereafter
we have  a region of saturation of the mean width and the separation
 point for the
second broad state. In fig. 5.c we see first the separation of 4 strongly
coupled states and the saturation of the widths up to the coupling strength
at which 10 states corresponding to the  10  channels  are
separated from the remaining ones. Finally the mean width of all the trapped
states decreases.\\

In the calculations  with the principal value integral ${\cal P}$
taken into account, we additionally see another effect (figs. 5.b,~d).
There are, at strong coupling to the continuum, not only $K$ broad states,
but more states separate from the
remaining ones and get large widths. In fig. 5.d we have at
$\kappa^{\rm tot}= 60$ about $15$ states
that are broad but the sharp distinction between the lifetimes of the
broad and the trapped states is washed out.\\

Similar results are obtained by studying other resonance states with
other quantum numbers.\\

In the eigenvalue picture, fig. 6.a, we show $\tilde E_R$
and ${{\tilde \Gamma}_R}$ for the same resonance states as in fig. 5.d
by varying $\kappa^{\rm tot}$ in the interval
$0.008 \le\kappa^{\rm tot}\le 60$. Note that the steps in $\kappa^{\rm tot}$
are approximatively equividistant. We see extra shifts
in energy caused by the principal value integral ${\cal{P}}$. Such a
shift is, generally, in the order of magnitude of the width of the state, i.e.
it is large for states getting large widths with increasing coupling
strength to the continuum.
That means, the broad states leave the energy region where the trapped states
are lying. These shifts are similar to those in figs. 4.a,~b for
$\theta\ne0$.\\

Fig. 6.a  illustrates also the behavior of trapped states under the
influence of increasing coupling to the continuum. First their widths
increase, then the states get trapped, i.e. their widths start to decrease and
they get a small energy shift. For even stronger coupling to the continuum
the widths can start to increase again with a renewed energy
shift and so on. In distinction to figs. 1.b and 3.c, these shifts
have mainly two origins: the energy attraction accompanying the
bifurcation of the widths {\it and} the influence of ${\cal P}$.\\

The corresponding $| \tilde \Phi_R|^2$  as a function of
$\kappa^{\rm tot}$ are shown in fig. 6.b.
The $| \tilde \Phi_R|^2$ remain larger than $1$ also
at large values of $\kappa^{\rm tot}$. This indicates further avoided
resonance crossings which cause, ultimately,
the saturation of the widths of the long-lived resonance states.\\

Summarizing, the $\cal P$ enlarges the critical region of reorganization
where the local process of resonance trapping takes place.
The mean width of the long-lived states saturates but there is
no longer a sharp distinction between the broad states and the long-lived
 ones.\\

\section{Resonance trapping and broadening of the width distribution}
\label{sec:width}

In the RMM with equally strongly coupled channels it is shown in
e.g. \cite{gorin,fyod} that the width distribution in the critical region of
reorganization is broader than in non-critical regions. In non-critical
regions, i.e.~as long as all transmission coefficients are small, 
the width distribution follows a $\chi_K^2$ law with the number of degrees
of freedom corresponding to the number $K$ of open channels
\cite{sokzel88}. In this section
we study the width distribution in the RMM with varying coupling strengths
to the channels and in the CSM.\\

In the RMM with varying coupling strengths of the channels we have tried
to fit the width distrubution in non-critical regions to a $\chi_K^2$
distribution. The error in the fit is large even for small transmission
coefficients. We
conclude that the system must be coupled to all the channels with comparable
strengths for the widths to be $\chi_K^2$ distributed.\\

In the following we therefore study the broadening of
the width distribution by calculating the normalized variance
$\sigma_y^j$ of the widths,
\begin{eqnarray}
\sigma_y^j=\sqrt{\frac{1}{N-j}\sum_{R=1}^{N-j}(y_R-1)^2} \; \; \; \; \; \;
y_R=\Gamma_R/{\bar\Gamma_{n-j}}
\label{eq:vargam}
\end{eqnarray}
where the sum runs over all but the $j$ broadest states.
The theoretical value obtained in the
RMM with $K$ equally strongly coupled channels
for the (trapped) states far from the critical region
of reorganization is $\sigma_y^{j=K}=\sqrt{2/K}\equiv\sigma^{\rm RMM}_y$.\\

The results of calculations in the RMM with varying coupling strengths to the
channels are
shown in fig. 7. In fig 7.a $\sigma_y^j$ is shown for $K=10$ channels
with $V_c^2/V_{c=1}^2$ distributed on $[1~...~ 0.1]$ with equal distances
at the logarithmic scale for $j=0...13$.
In figs. 7.b and c we show $\sigma_y^j$ for $K=4$ channels and $j=1...6$ with
${\rm V_{\rm c}^2}/{\rm V_{\rm c=1}^2}=1$, $0.1$, $0.01$, $0.001$ and
${\rm V_{\rm c}^2}/{\rm V_{\rm c=1}^2}=1$, $0.01$, $0.0032$, $0.001$,
respectively (compare figs. 3.a,~b). Note that $\sigma_y^j$ before
the separation of a broad state should be compared with the
value $\sigma_y^{j+1}$ after the separation.\\

In the figures, the values $\sigma_y^{\rm RMM}=\sqrt{2/K}$ are shown with
a dashed
line. In the case of different coupling strengths of the $K$ channels,
the width distribution is broader than in the case of $K$ channels
with comparable coupling strengths.\\

The separation of every broad state is accompanied by a broadening of
the width distribution, see figs. 7.a to c. The distribution
between the separation points  in fig. 7.b and between the separation points
of the first and second broad state in fig. 7.c. is however narrower than the
distributions for very small and very large $\kappa^{\rm tot}$.\\

To describe the distribution of the long-lived states in regions
where all $\tau_c<1$ we make the following ansatz,
compare Eq. (\ref{eq:mosi2}):
\begin{eqnarray}
\Gamma_R^l=-\frac{d}{2\pi}\sum_c \ln(1-\tau_c) ~ g_{R,c}^2 \; .
\label{eq:granz}
\end{eqnarray}
Here $g_{R,c}^2$ for a certain $c$ is a Gaussian distributed vector with
mean zero and variance $1$ and $\tau_c$ is the transmission coefficient
defined in Eq. (\ref{eq:mosi1}).\\

The normalized variance of $\Gamma_R^l$ is
\begin{eqnarray}
\sigma_y^l=
\frac{\sqrt{\overline{(\Gamma_R^l)^2}-\left(\overline{\Gamma_R^l}\right)^2}}
{\overline{\Gamma_R^l}}=
\frac{\sqrt{2\sum_c~\left(\ln(1-\tau_c)\right)^2}}{-\sum_c~\ln(1-\tau_c)}
\geq \sigma_y^{\rm RMM}\; .
\label{eq:varanz}
\end{eqnarray}
(Note that $\overline{g^2}=1$ and $\overline{g^4}=3$.)
$\sigma_y^l=\sigma_y^{\rm RMM}$ for the same number $K$ of open decay
channels holds only when all $\tau_c$ are equal.
The values calculated from (\ref{eq:varanz}) are shown with a thick line
in figs. 7.a to c. The variances are well described by (\ref{eq:varanz})
in regions of the coupling strength $\kappa^{\rm tot}$ with all $\tau_c<1$ .\\

The values $\sigma_y^j$ as a function of $\kappa$ calculated in the CSM are
similar to those obtained in the RMM. One example  is shown in fig. 7.d
for the case $J^\pi=2^-$ with $K=12$ open channels.\\

We conclude that in many-body quantum systems, the distribution of the
widths is broader than in the  RMM with equal
coupling strength to the channels (also at small transmission coefficients).
This result explains the broadening of the width distribution described in
the literature by introducing an effective number $K^{\rm eff}$
of channels being smaller than $K$.\\

For the case shown in fig.~7.a, we have calculated also
 $k^{\rm eff}_{\rm gr}$, Eq. (\ref{eq:rate6}), as a
function of time $t$  with $12$ different coupling
strengths $\kappa^{\rm tot}$ of the system to the channels
(fig.~8). $\hbar=1$.
 In table \ref{tab:keff} the values $\kappa^{\rm tot}$
and $\bar\Gamma_{N-K}$ for the different curves are given.
All the curves a to f, being below the critical region,
differ considerably from one another. Inside the critical region, the
curves for different $\kappa^{\rm tot}$ (g to i)
are, however, similar to one another.  \\

As a result, 
weakly coupled channels  cause
 a saturation of the mean value $\bar\Gamma_{N-j}$
as a function of $\kappa^{\rm tot}$. Since the width distribution
$\sigma_y^j$ does not change much,  also $k^{\rm eff}_{\rm gr}$
saturates in the long-time scale, i.e. $k^{\rm eff}_{\rm gr}$ remains
almost unchanged by varying $\kappa^{\rm tot}$ in the critical region.
The saturation is related to the enlarged width distribution.\\

\section{Conclusion}
\label{sec:concl}

In this paper we studied the positions and widths of
resonance states in a many-body quantum  system with
 $N$ resonance states as a function of increasing coupling strength
$\kappa^{\rm tot}$ to the continuum which consists of $K \ll N$ decay channels.
In a critical region of the coupling strength, the
system reorganizes itself under the influence of the decay channels. The
local process is the avoided crossing of two resonance
states which takes place
whenever the distance in energy between the states is comparable to the
sum of their widths. It is accompanied by an essential
bi-orthogonality of the eigenfunctions of  $H^{\rm eff}$. As a result,
one state continues to increase in width whereas the other one decreases
with further increasing coupling strength (resonance trapping). For a system
with  many states, this leads to a broadening of the width distribution.\\

With further increased coupling strength, the broadening of the width
distribution goes over into a separation of time scales if $K \ll N$.
The smaller $K$ the better expressed is this separation.
Weakly coupled channels as well as  the hermitean part of
the coupling term $\hat V_{\rm QP} G_{\rm P}^{(+)} \hat V_{\rm PQ}$
have the tendency of washing out the
differences between the lifetimes of the group of long-lived trapped states,
on the one hand, and the group of short-lived states, on the other hand.\\

We studied in detail the widths of the $N-j$ long-lived trapped states
under different
conditions. We introduced additional channels to which the system is weakly
coupled and we took into account the hermitean part of
${\hat V}_{\rm QP} G_{\rm P}^{(+)}{\hat V}_{\rm PQ} $ in
$H_{\rm QQ}^{\rm eff}$, Eq.
($\ref{eq:heff}$). Under the influence of these additional
terms in the Hamiltonian, the trapped states can increase their
widths and change their positions in energy. Thus, trapped resonance
states
may again come close to each other if the coupling to the continuum is
stronger. The resonance crossing is avoided and accompanied by an essential
bi-orthogonality of the eigenfunctions of  $H^{\rm eff}$
in the same manner as at smaller coupling strength. As a result
of all these processes,  the average width of the states
saturates as a function of the coupling strength to the continuum
when an appropriate number of states with the largest widths
is excluded from the mean value.
This number is equal to the number of open decay channels as long as the
real part of the coupling term
$\hat V_{\rm QP} G_{\rm P}^{(+)} \hat V_{\rm PQ}$ is small compared to
its imaginary part.\\

The width distribution of the long-lived states is related to the transmission
coefficients. If all transmission coefficients are equal and smaller than 
one, the
widths are $\chi^2_K$ distributed where $K$ is the number of
channels. In many-body quantum systems with different
coupling strengths to the different decay channels, the distribution is
broader than a $\chi^2_K$ distribution also for small 
transmission coefficients.\\

The decay rates are related to the mean decay widths of the long-lived states.
This means,  also the decay rates saturate  in quantum systems at high
level density as a function of the coupling strength $\kappa^{\rm tot}$. \\

Summarizing the results we state that
resonance trapping is a realistic process occuring in many-particle quantum 
systems at high level density. It leads to  a saturation
of both the decay rates and the average decay width of the long-lived
states as well as to a broadening of the
width distribution and -- if the number of open decay channels is not
too large -- to a separation of time scales. That means, the decay rates and
the decay widths of the long-lived resonance states show the same
behavior not only at low level density but also at high level density. The
saturation is caused in both cases
by the finite number $K < N$ of channels into which the
$N$ resonance states can decay. The application of the standard random matrix
approach to the details of the trapping process in an ensemble of
resonance states is limited.\\

We would like to state once more  that
the time de-excitation of resonance states at high level density should be
directly measured. The results could make   a proof of the
phenomenon of resonance trapping possible.\\

{\small
\noindent
{\bf Acknowledgment:} We gratefully acknowledge  valuable discussions with
 V.A. Mandelshtam, M.~M\"uller, H.~Reisler, T.H.~Seligman,
G.~Soff and H.S.~Taylor.  This work is supported by DFG (Ro
922).}\\

\newpage

{\bf Figure 1}\\

\noindent
$\kappa^j$ for $j=0,1,2$ and $2 \langle \Gamma_l \rangle /D_0$,
calculated from Eq. ($\ref{eq:mosi2}$) (thick line),
versus $\kappa^{\rm tot}$ {\bf (1.a)}. Eigenvalue picture
($\frac{1}{2}\Gamma_R$ and $E_R$) calculated for different
$\kappa^{\rm tot}$  {\bf (1.b)}.
The calculation shown is performed within the RMM for $K=1$ and $N=300$.
In (1.a) the curves shown are averages over $20$ calculations. Only
a part of the spectrum is shown in (1.b). Note the logarithmic scales.
The points for $\kappa^{\rm tot}=0.1$, $1$ and $10$
are marked  in (1.b) with triangles, stars and squares, respectively.\\

\vspace{1cm}

{\bf Figure 2}\\

\noindent
Eigenvalue picture
($\frac{1}{2}\Gamma_R$ and $E_R$) for $H_{\rm b}$ consisting of
a sum of three Gaussian shapes. $N = 300, \; K = 4$ and all
channels are coupled with the same strength.\\

\vspace{1cm}

{\bf Figure 3}\\

\noindent
$\kappa^j$ for $j=0...6$ versus $\kappa^{\rm tot}$ in the RMM for $N=300$
and $K=4$ with the ratios of the coupling strengths
${\rm V_{\rm c}^2/\rm V_{\rm c=1}^2}=1$, $0.1$, $0.01$,
$0.001\;$ {\bf (3.a)} and
${\rm V_{\rm c}^2/\rm V_{\rm c=1}^2}=1$, $0.01$, $0.0032$, $0.001$
 {\bf (3.b)}.
Eigenvalue picture
{\bf (3.c)} and
 $\frac{1}{2}\Gamma_R$ (dots) together with the function $
\frac{1}{2} (x_c - 1/x_c)\; $,
Eq.~($\ref{eq:wshort}$), (full line)  versus 
$\kappa^{\rm tot}$ 
{\bf (3.d)} for ${\rm V_{\rm c}^2/\rm V_{\rm c=1}^2}=1$, $0.01$,
$0.0032$ and $0.001$. 
In (3.a) and (3.b), $2 \langle \Gamma_l \rangle /D_0$ versus
$\kappa^{\rm tot}$ is shown with a thick line.
In (3.c), the points for
$\kappa^{\rm tot} = 0.1$, $1$, $10$, $100$ and $1000$ are marked   with
triangles, stars, diamonds, large dots and squares, respectively.
\\

\vspace{1cm}

{\bf Figure 4}\\

\noindent
Eigenvalue picture ($\frac{1}{2}\Gamma_R$ and $E_R$)  calculated for
different $\kappa$ {\bf (4.a,~b)} and
$|\Phi|^2$ versus $\kappa$ {\bf (4.c,~d)} for two states coupled
to one channel. It is $\varphi=45^0$ (4.a,~c) and $\varphi=22.5^0$ (4.b,~d).
The values are shown for some different $\theta$ in the same plots
(see the text for details). Note the logarithmic scales.\\

\vspace{1cm}

{\bf Figure 5}\\

\noindent
$\kappa^j$ versus $\kappa^{\rm tot}$ in the CSM for $190$
resonance states in $^{16}O$ with $J^\pi=1^-$, $K=2$
{\bf (5.a,~b)} and $K=10\;$ {\bf (5.c,~d)} channels. In (5.a,~c)
${\cal P}=0$ while ${\cal P}$ is taken into account in (5.b,~d).
It is $j=0...4$ (5.a,~b), $j=0...14$ (5.c) and $j=0...25$ (5.d).
The calculations are performed at $E=29 $ MeV.\\

\vspace{1cm}

{\bf Figure 6}\\

\noindent
Eigenvalue picture
($\frac{1}{2}\tilde\Gamma_R$ and $\tilde E_R$) for different coupling
strengths $0.008 \le \kappa^{\rm tot} \le 60 \;$ {\bf (6.a)} and
$| \tilde \Phi_R |^2$ versus $\kappa^{\rm tot}\;$ {\bf (6.b)} in the CSM.
$N=190$ resonance states in $^{16}O$ with $J^\pi=1^-$ and $K=10$.
${\cal P}$ is taken into account (compare fig. 5.d).\\

\vspace{1cm}

{\bf Figure 7}\\

\noindent
$\sigma_y^j$, $\sigma_y^l$ and $\sigma_y^{\rm RMM}$
versus $\kappa^{\rm tot}$ in the RMM
for three selected values of $\rm V_{\rm c}^2/\rm V_{\rm c=1}^2$
(for details see the text). $\sigma_y^j$ is shown for $j=0...14$ {\bf (7.a)}
and $j=0...6$ {\bf (7.b,~c).} The thick lines
are $\sigma_y^l$ obtained from Eq. (\ref{eq:varanz}) and the dashed
lines are $\sigma_y^{RMM}$.
 {\bf (7.d)}:
$~\sigma_y^j$ for $j=1..20$ versus $\kappa^{\rm tot}$ in the CSM
for $J^\pi=2^-$,  $K=12$ and $\cal P$ is taken into account.
$j=0...20$ and the dashed line is $\sigma_y^{\rm RMM}$ for $K=12$. \\

\vspace{1cm}

{\bf Figure 8}\\

\noindent
$k_{\rm gr}^{\rm eff}$ versus time $t$ for $K=10$ and
$\rm V_{\rm c}^2/\rm V_{\rm c=1}^2=1...0.1$
(compare fig 7.a). The $12$ curves are calculated with $\kappa^{\rm tot}$
between $0.1$ and $50$. $\kappa^{\rm tot}$ and $\bar\Gamma_{N-K}$ for the
curves are given in table  \ref{tab:keff}.\\

\vspace{1cm}

\begin{table} [hp]
\caption{ $\kappa^{\rm tot}$ and $\bar\Gamma_{N-K}$
for the different curves  in fig. 8}
\vspace*{0.5cm}
\label{tab:keff}
\begin{center}
\begin{tabular}{|cll|}
\hline
 &  & \\
\hspace*{.7cm}
Curve
\hspace*{.5cm}
 &
\hspace*{.2cm}
  $\kappa^{\rm tot}$
\hspace*{.5cm}
  &
\hspace*{.1cm}
   $\bar\Gamma_{N-K}$
\hspace*{.7cm}
    \\[-.2cm]

   &  &  \\
 \hline
   & &  \\
a  &  ~0.10 & 0.094 \\
b  &  ~0.18 & 0.17  \\
c  &  ~0.31 & 0.29  \\
d  &  ~0.54 & 0.51  \\
e  &  ~0.96 & 0.89  \\
f  &  ~1.7  & 1.5   \\
g  &  ~3.0  & 2.6   \\
h  &  ~5.2  & 3.9   \\
i  &  ~9.2  & 4.7   \\
j  & 16.    & 4.9   \\
k  & 28.    & 4.5   \\
l  & 50.    & 3.3   \\
   & &  \\
 \hline
   \end{tabular}
    \end{center}
\end{table}

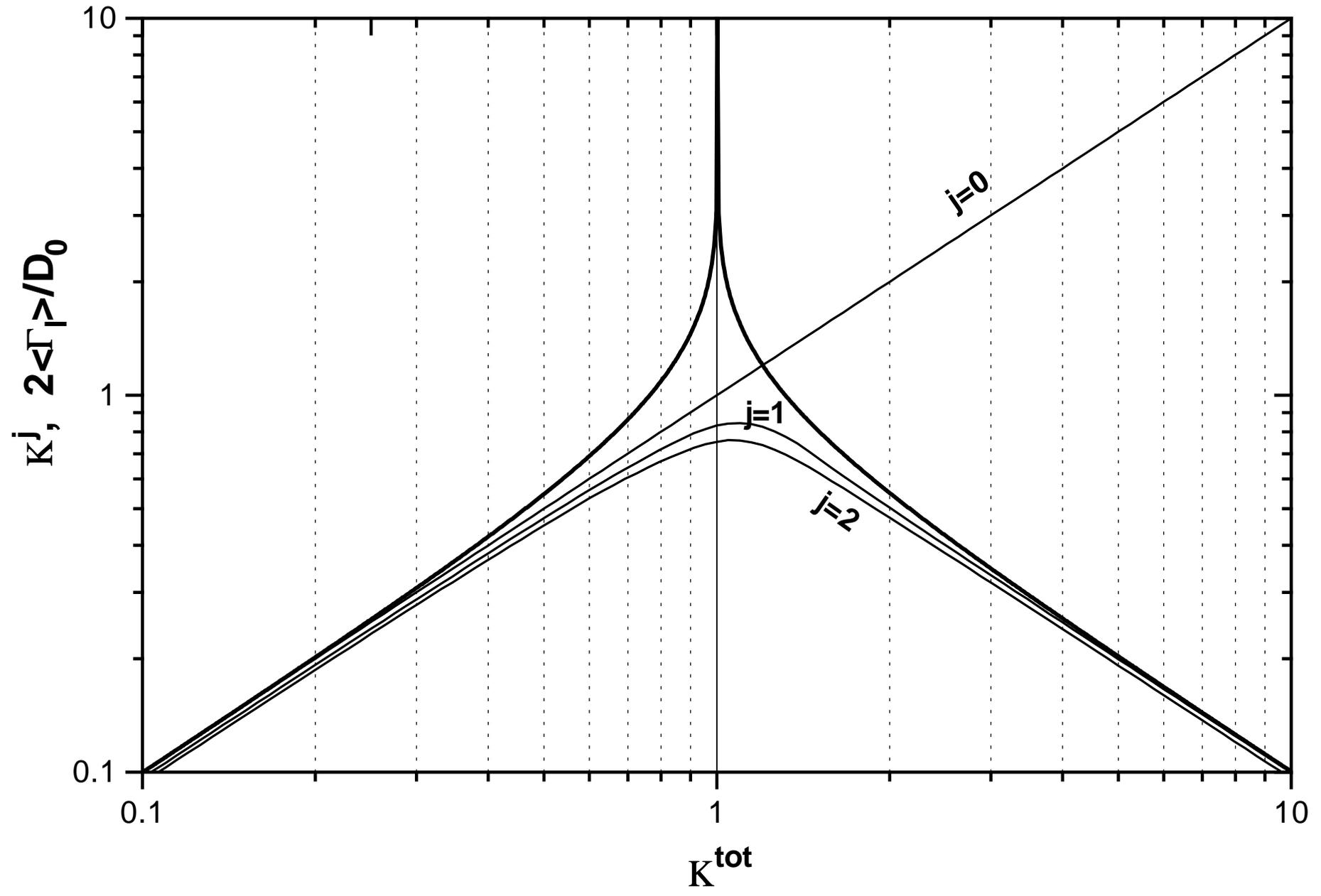
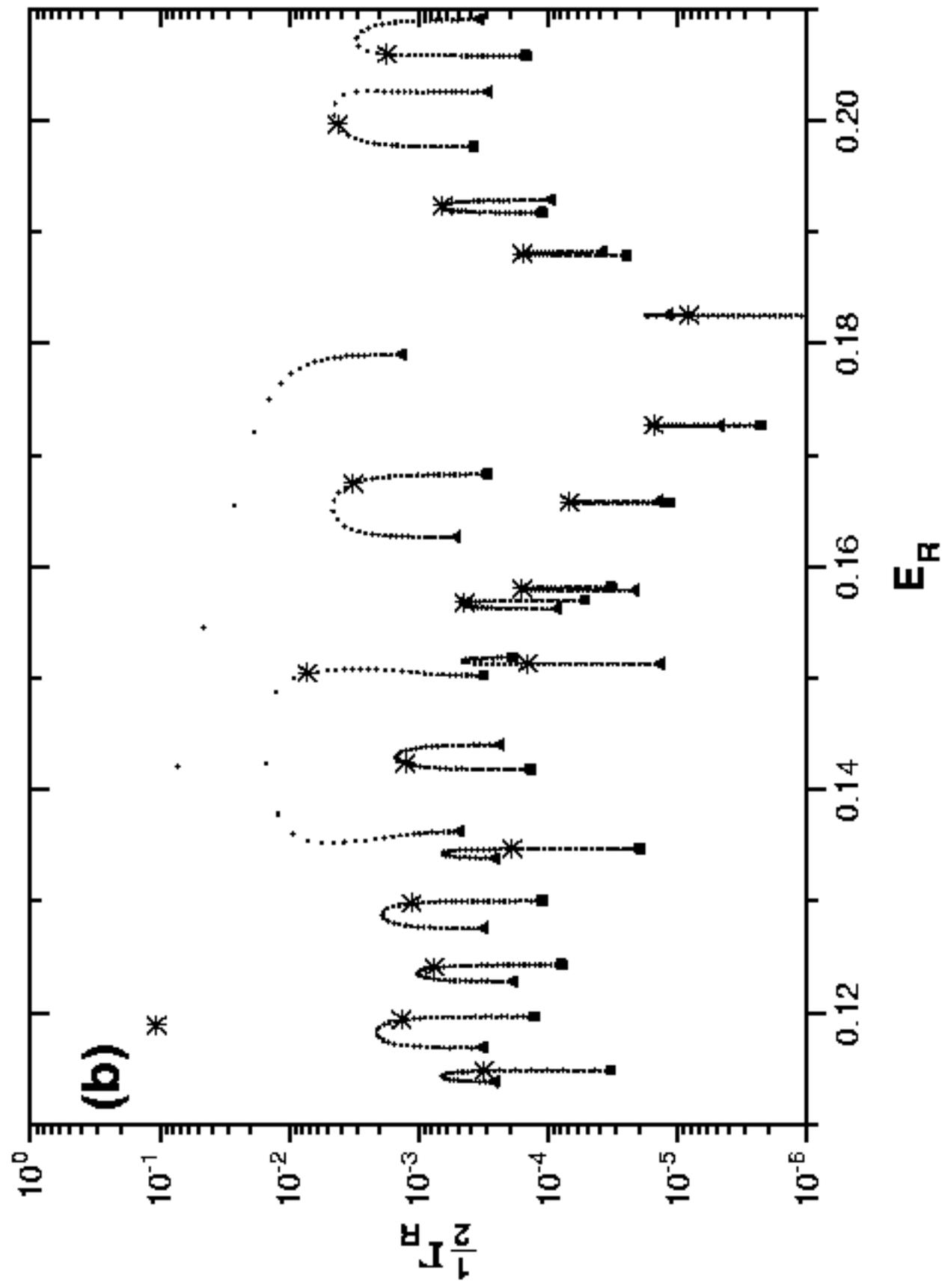
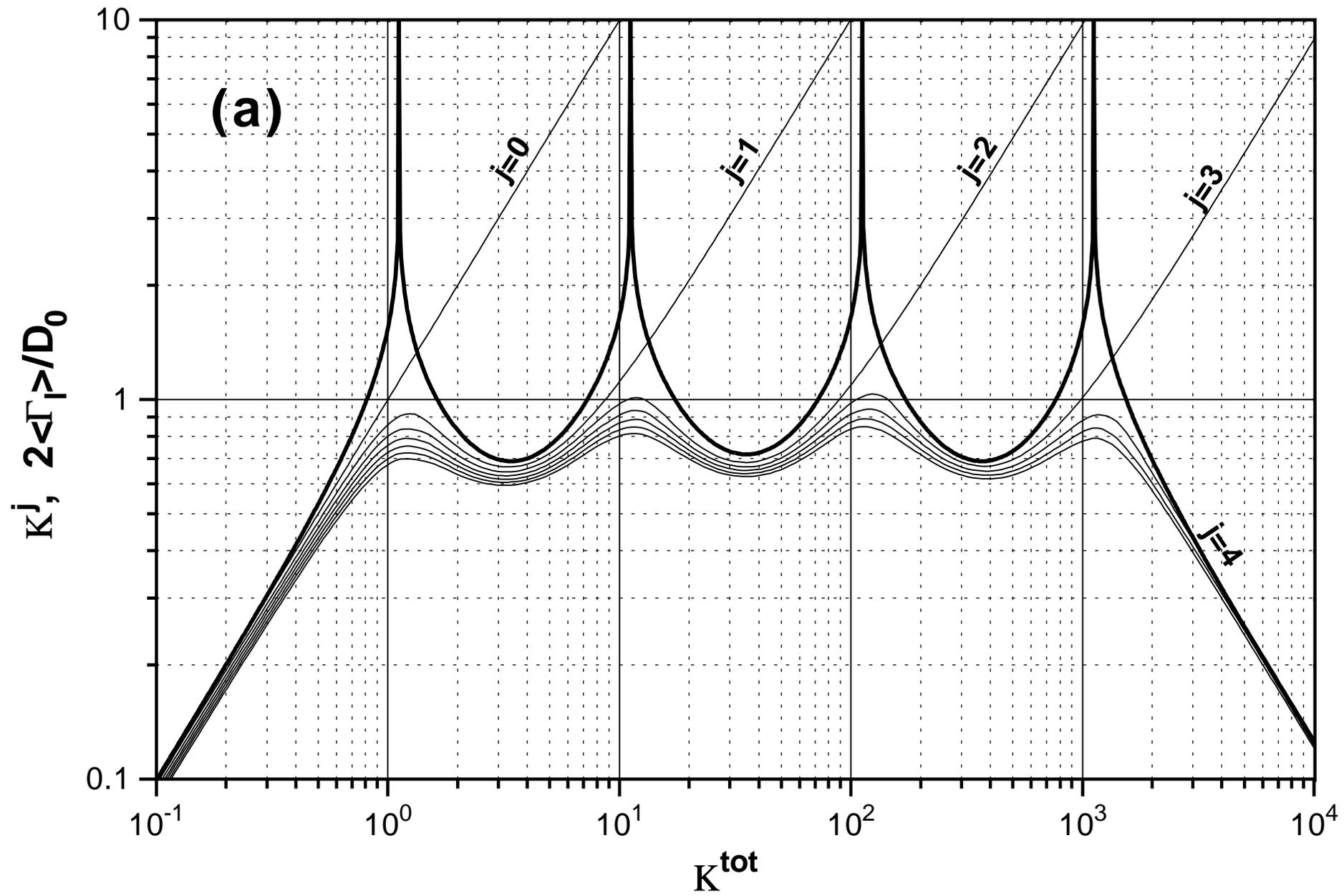
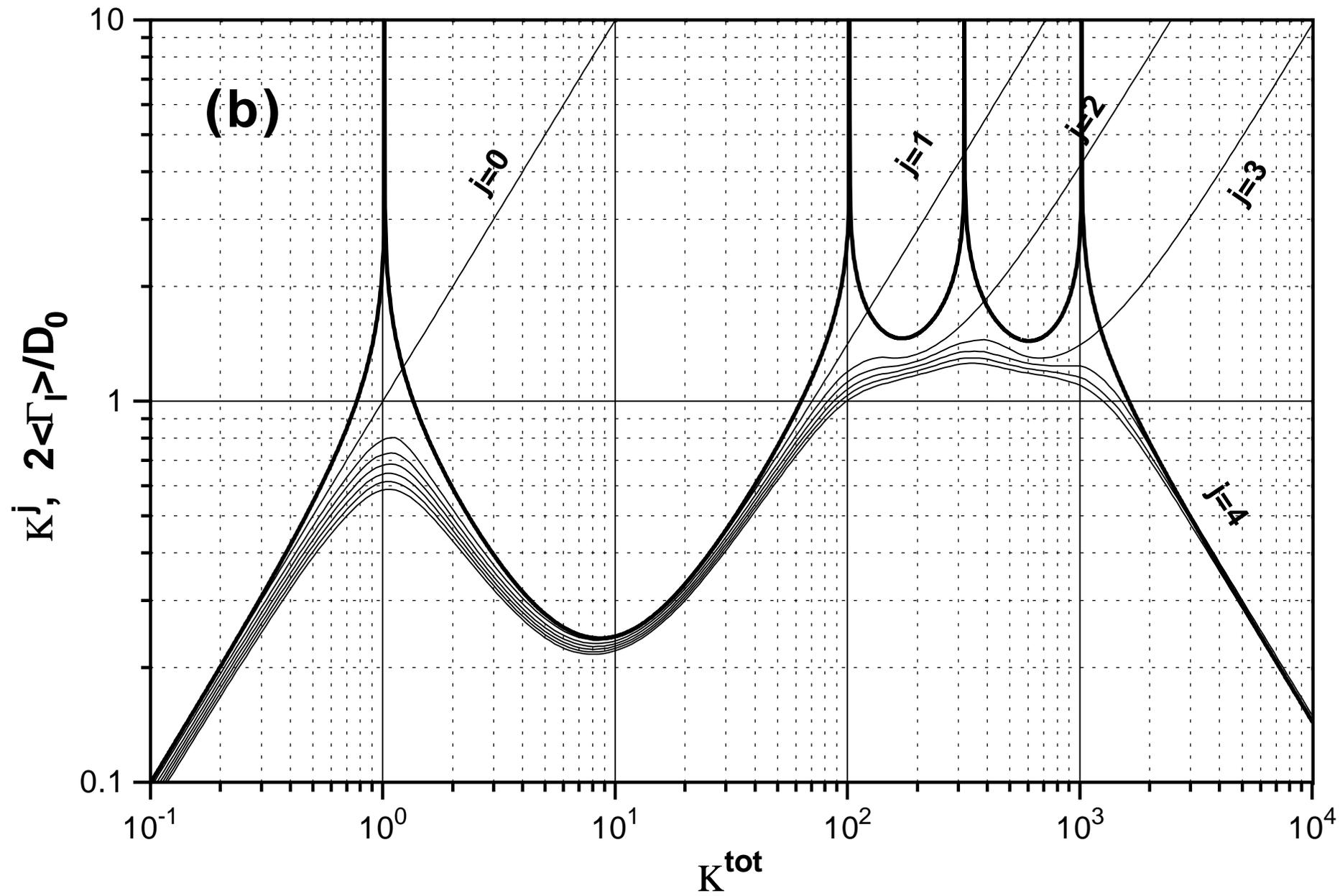
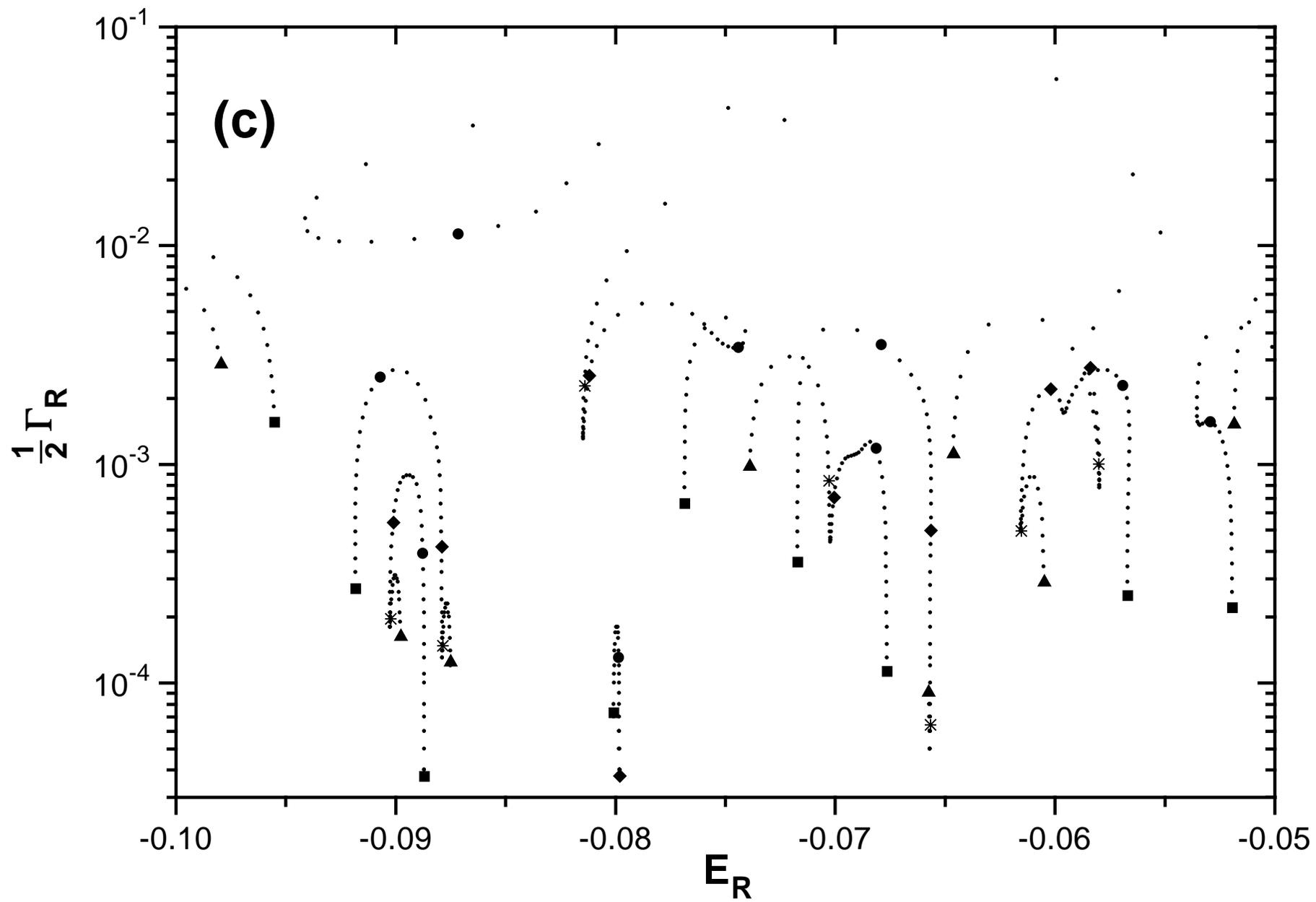
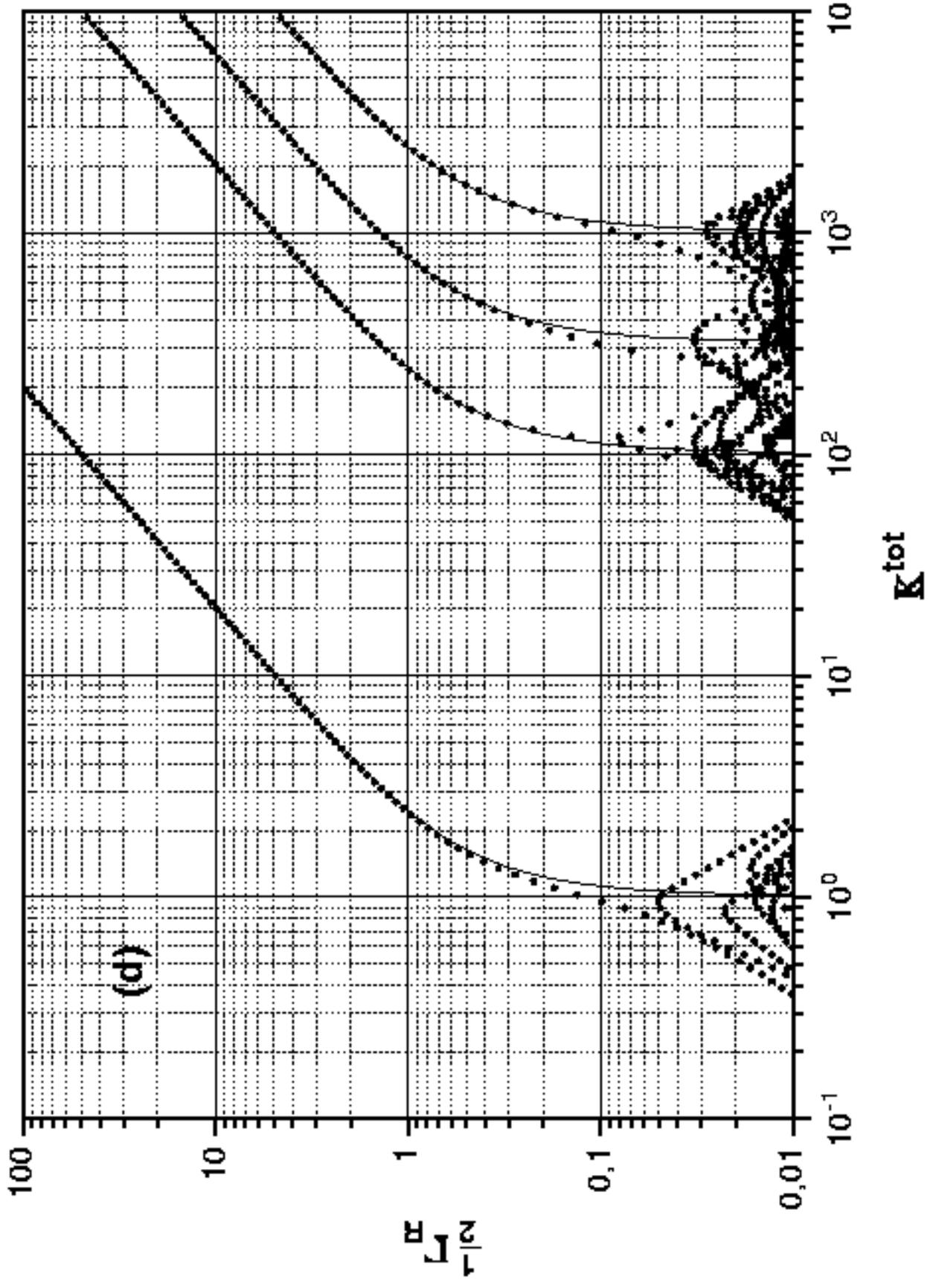
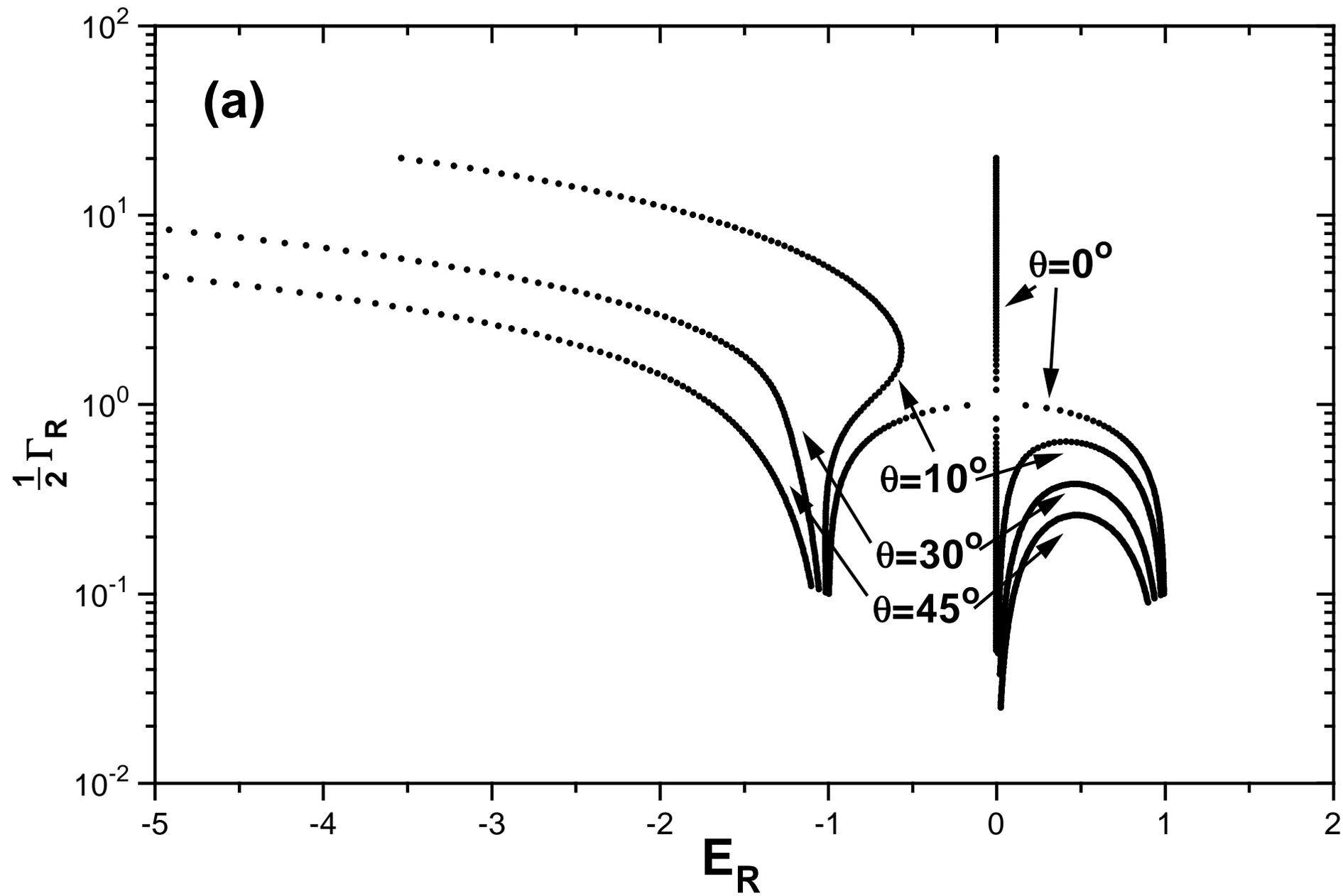
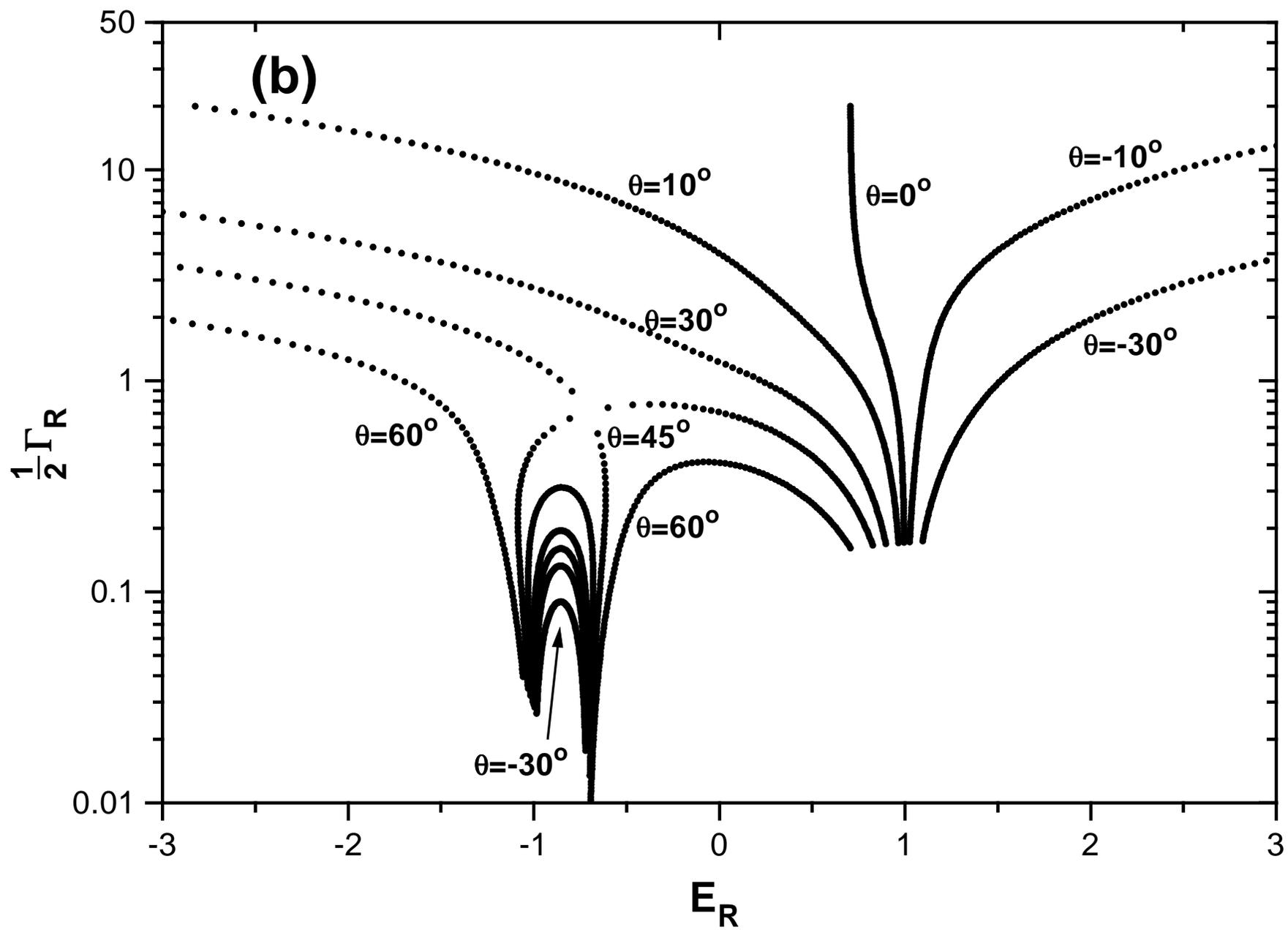
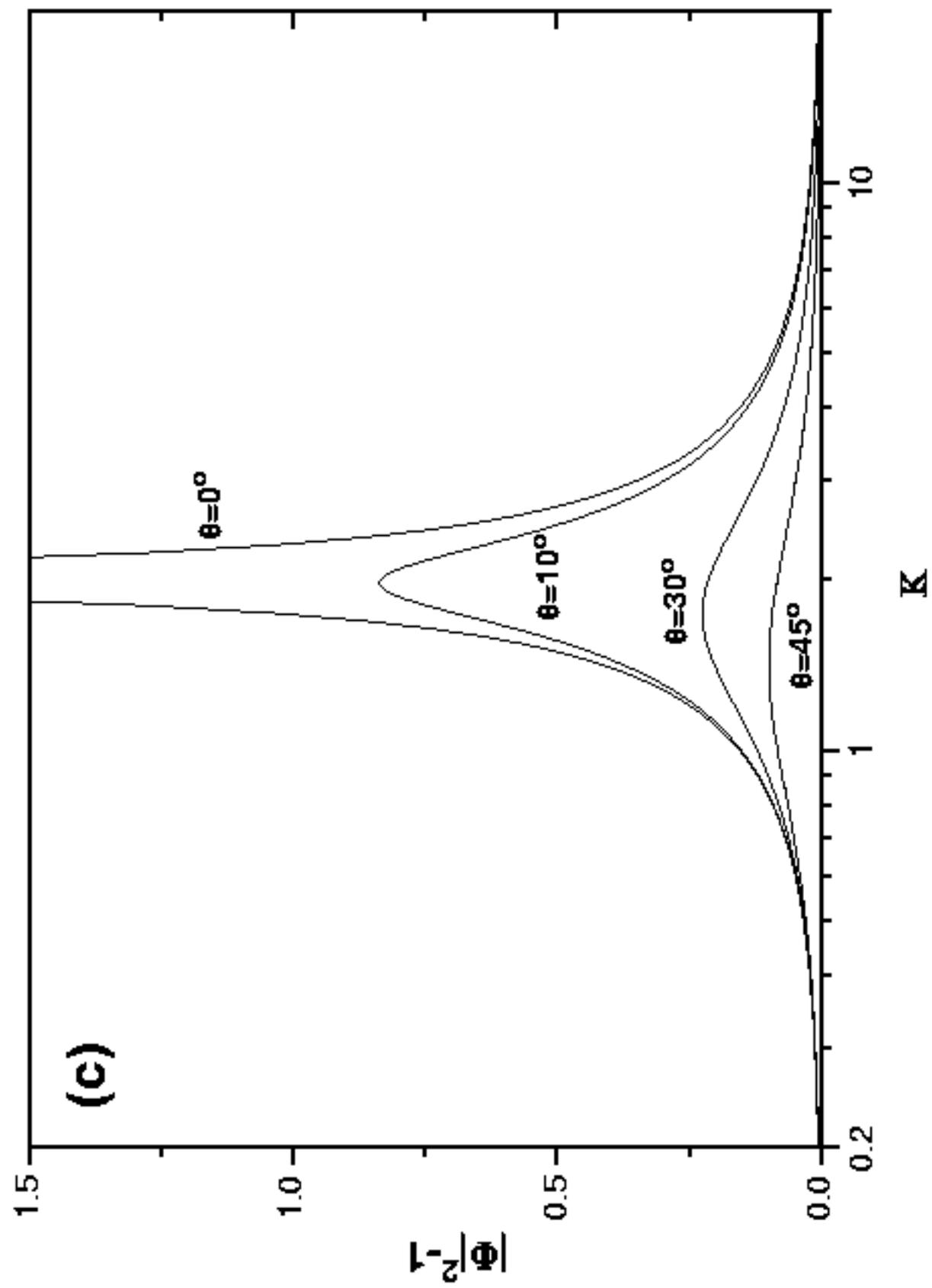
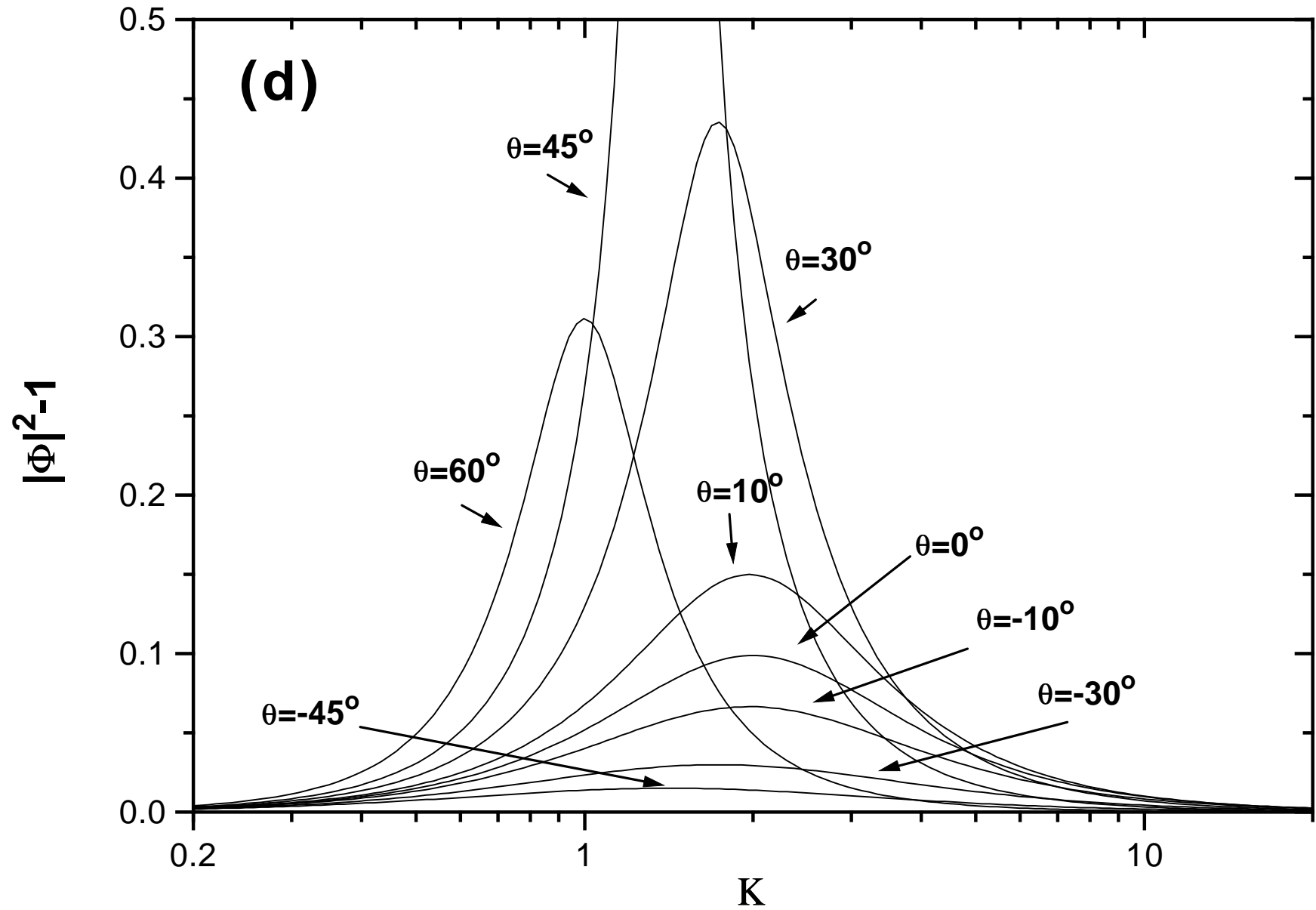
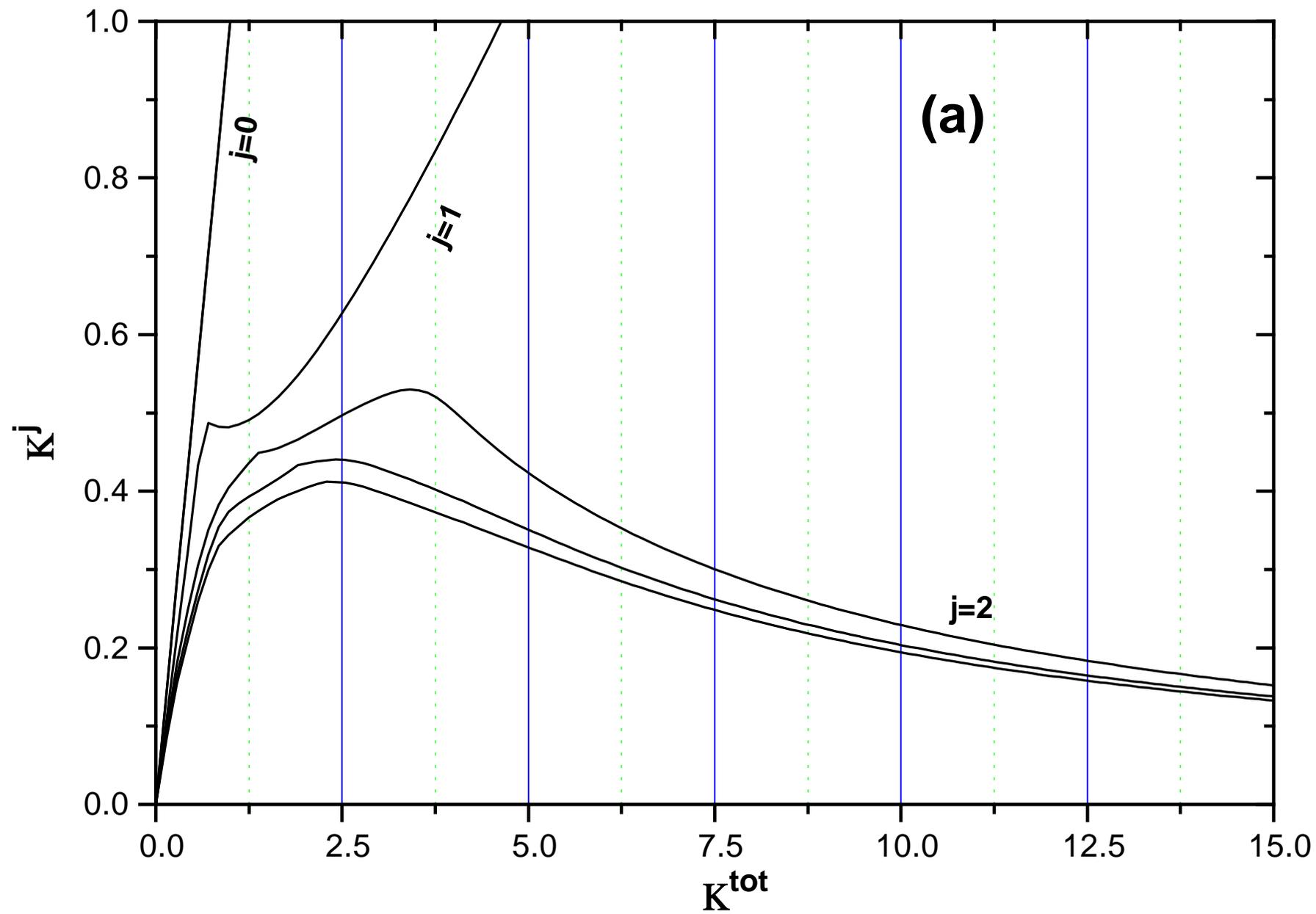
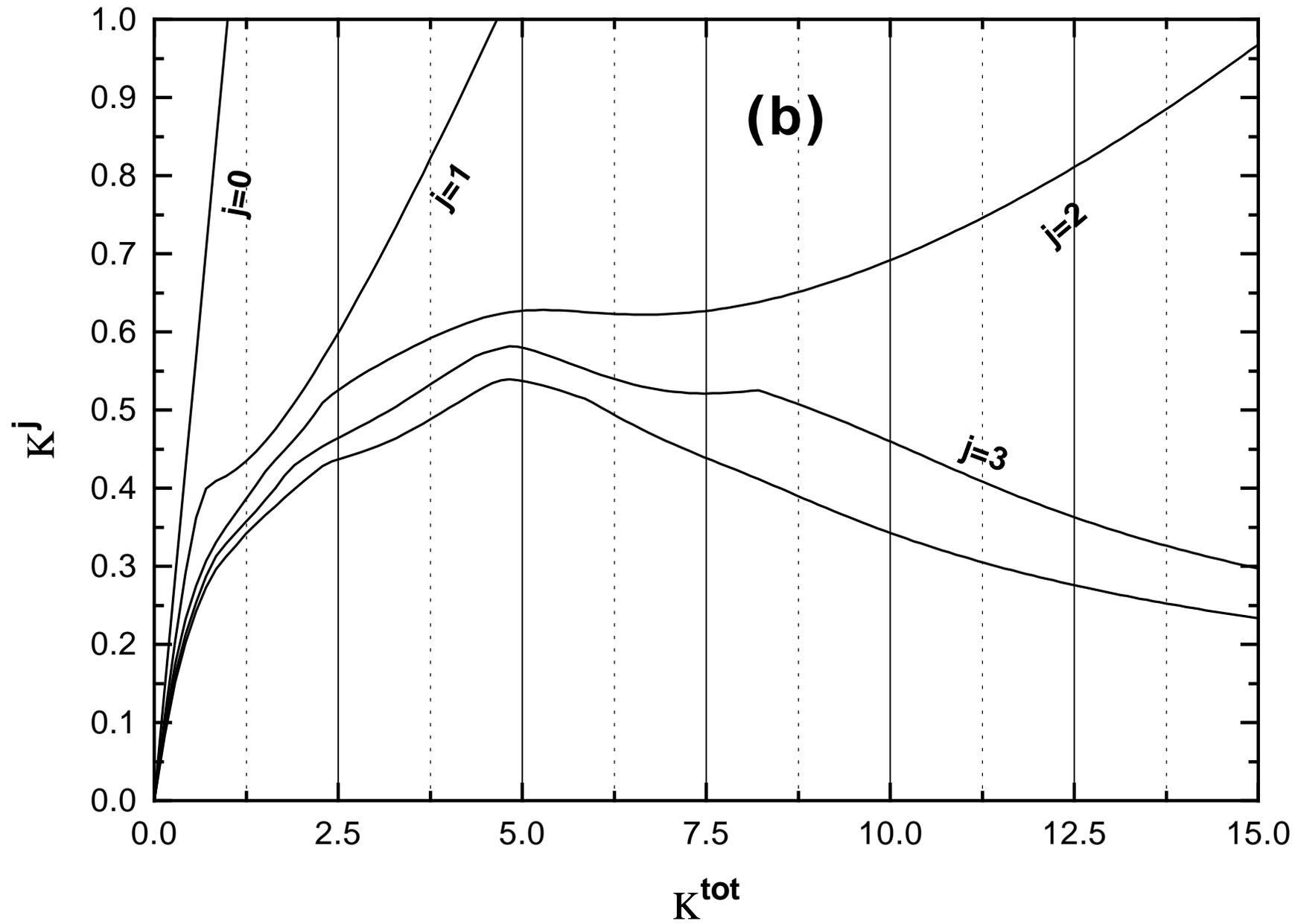
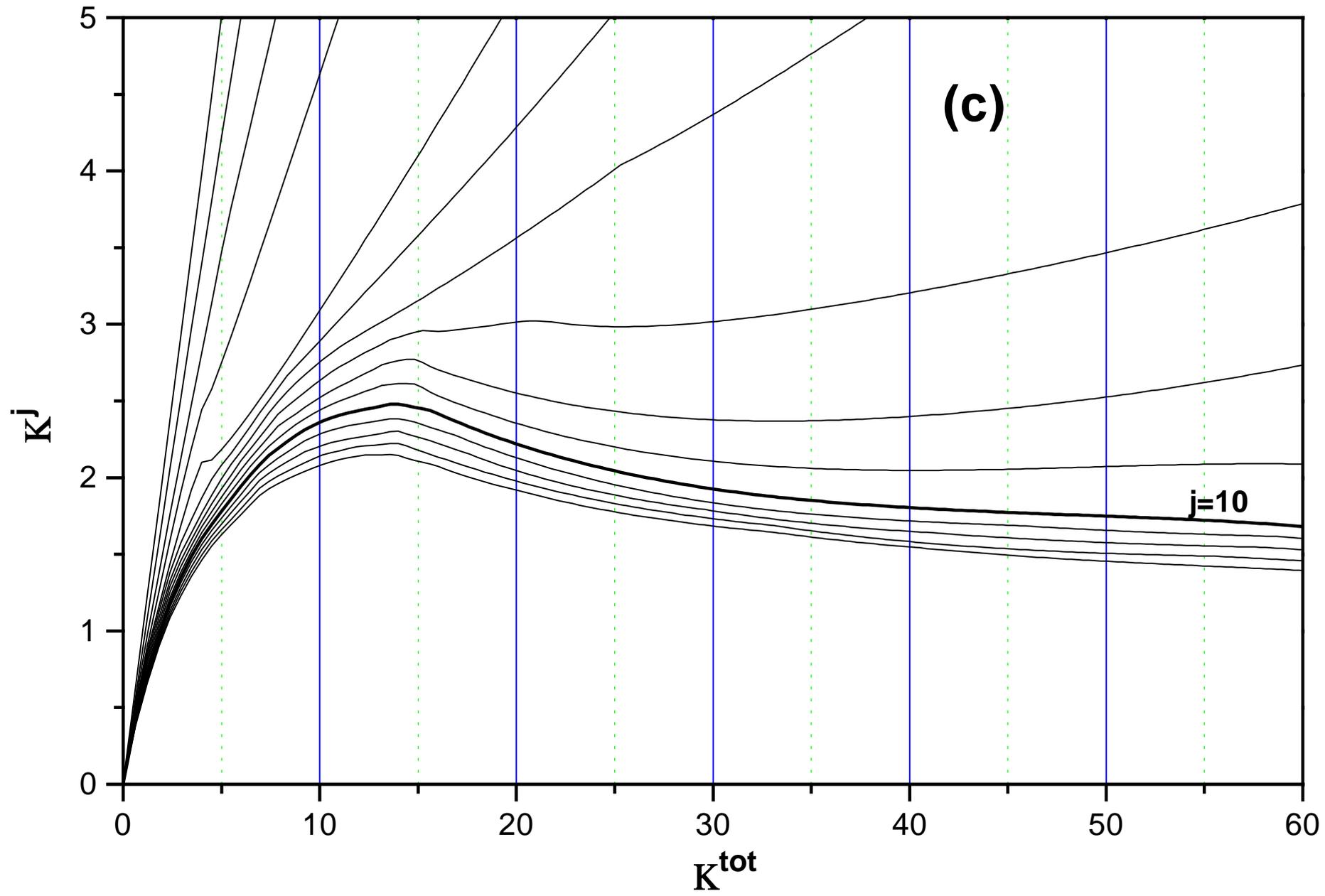
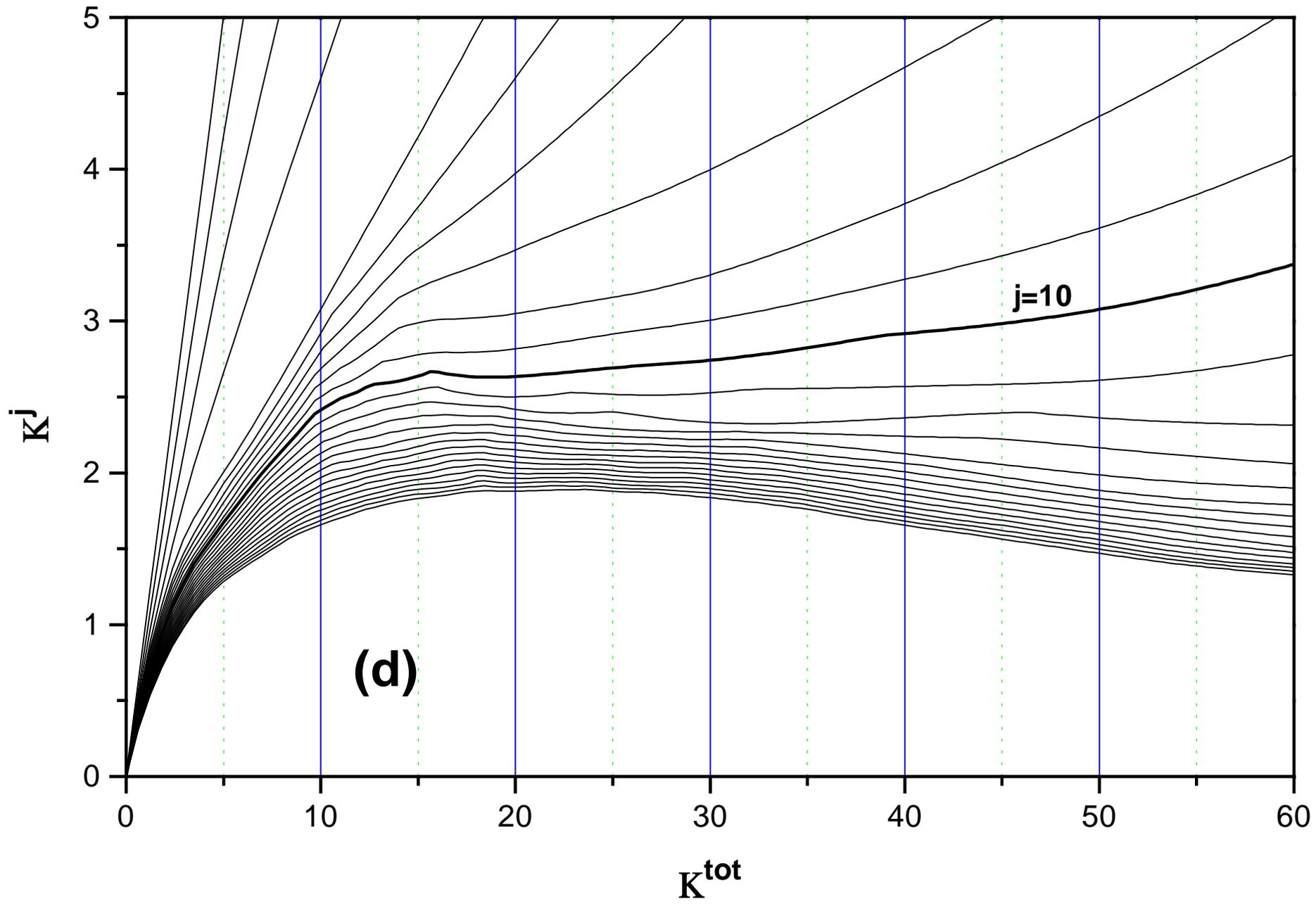
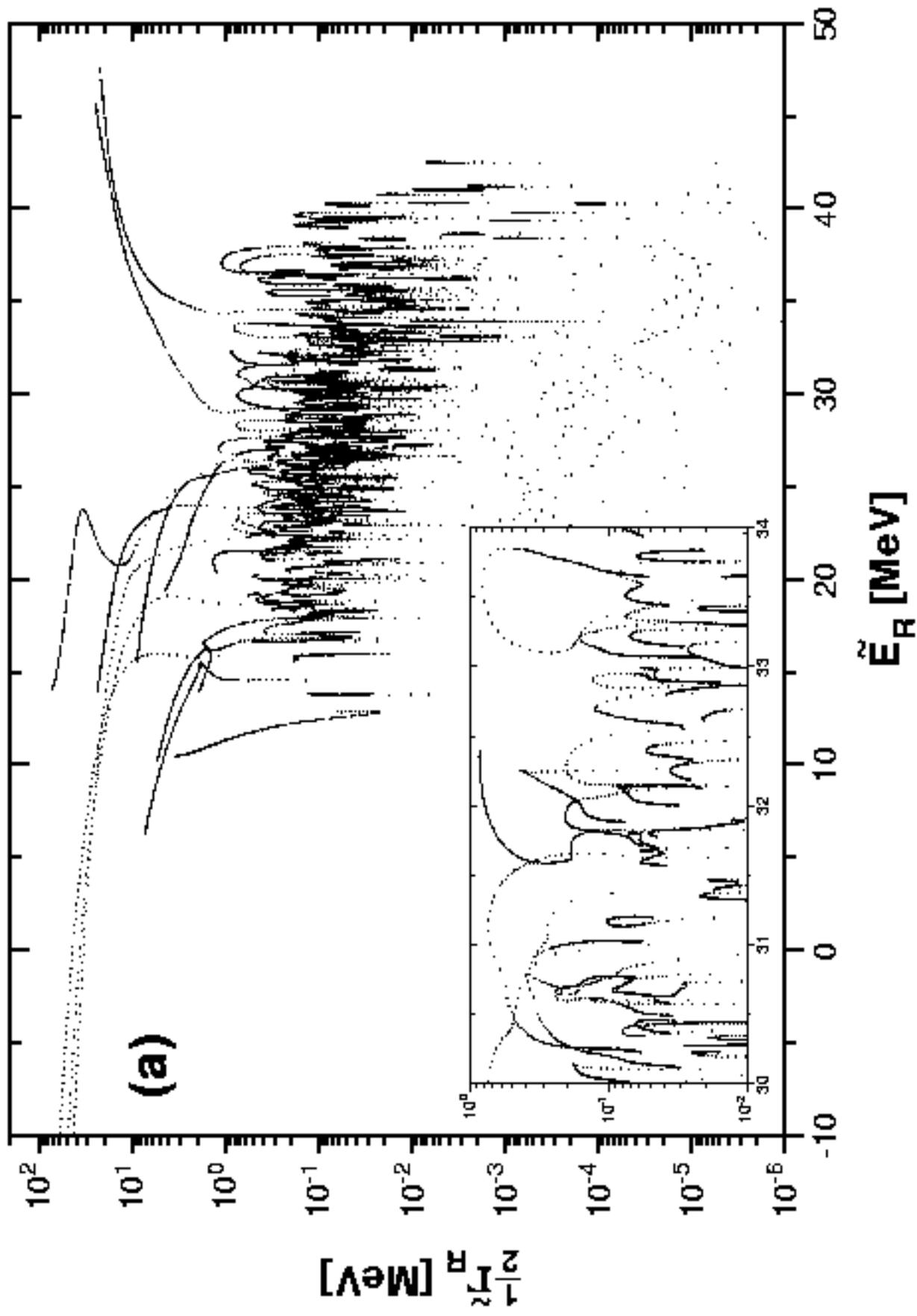
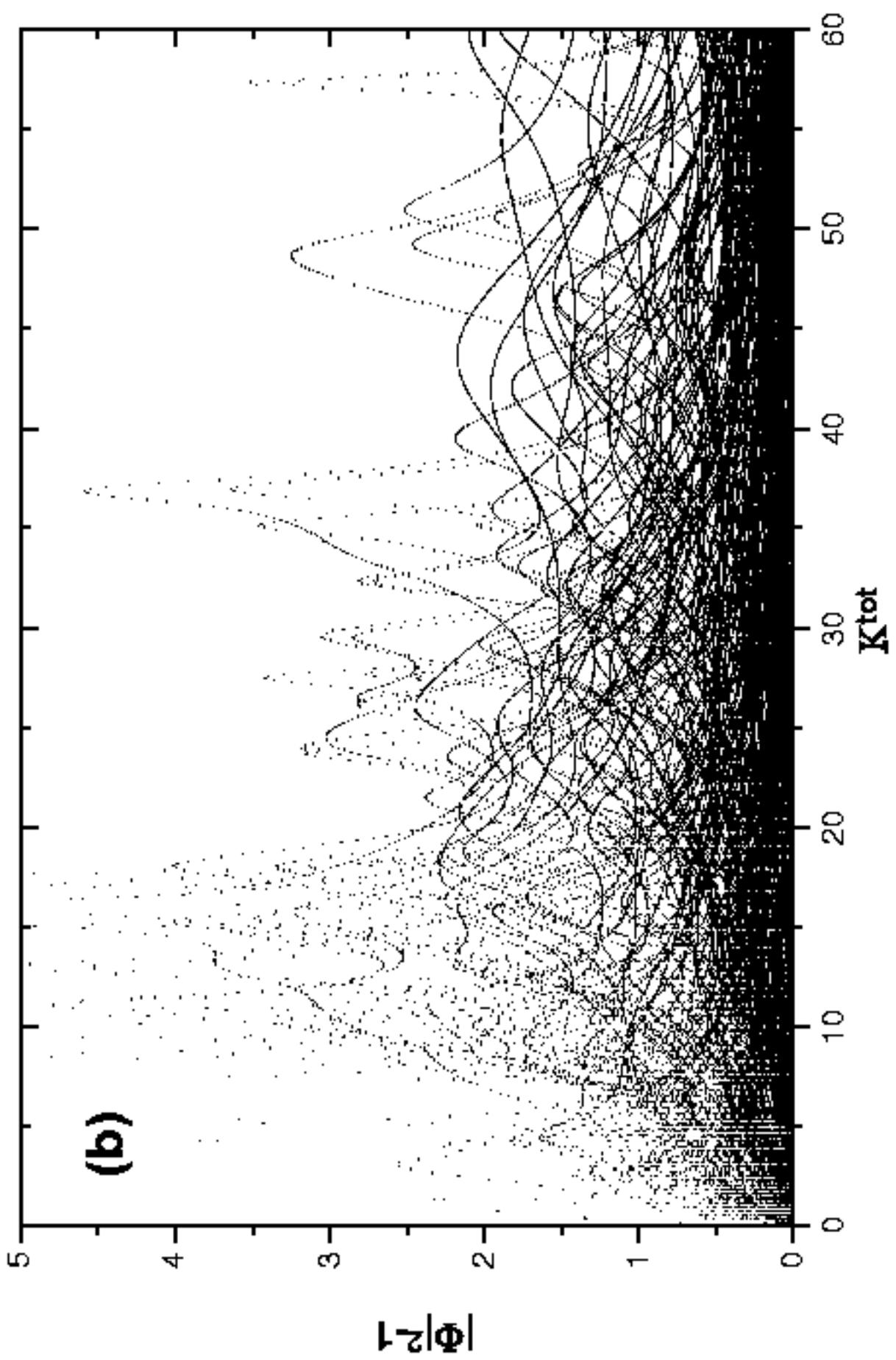
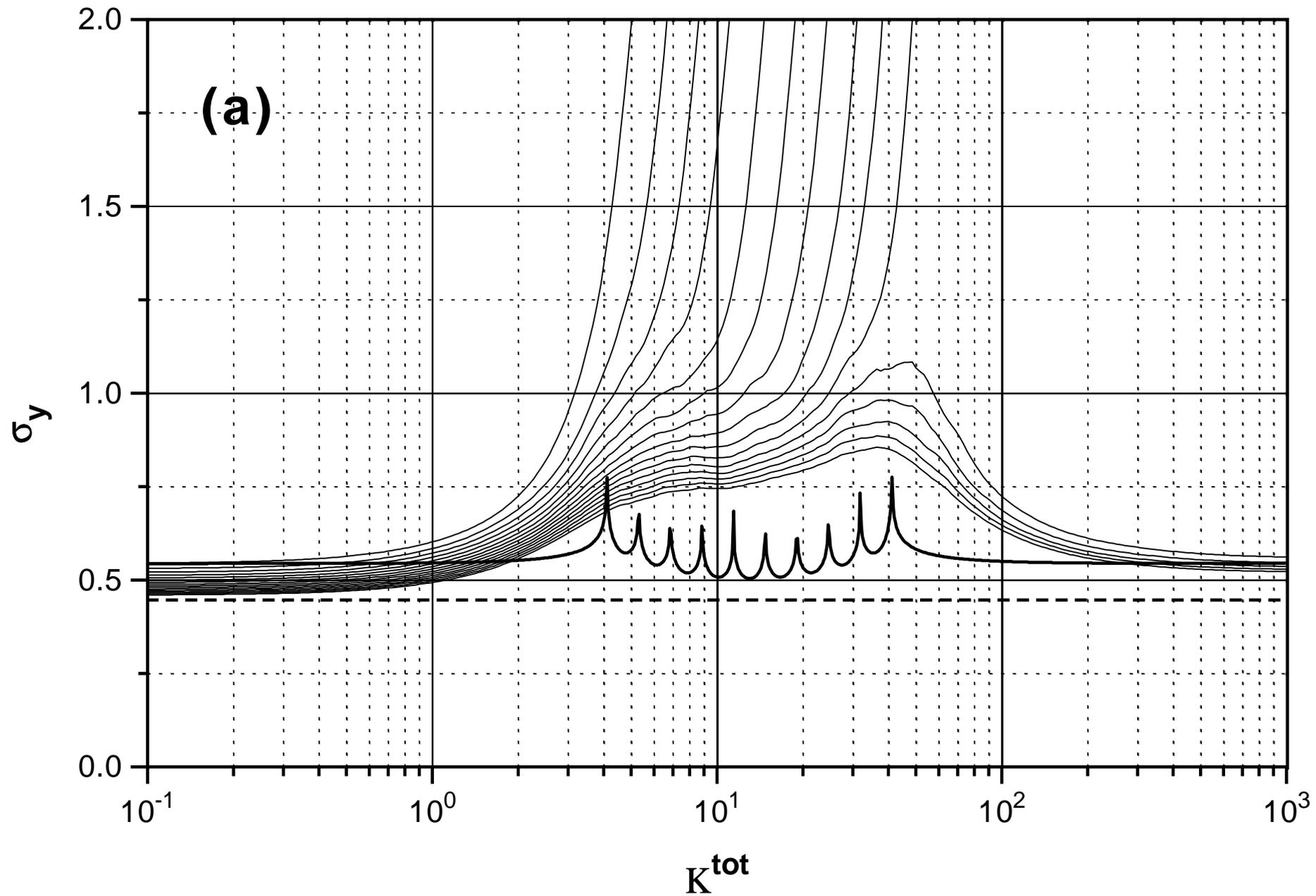
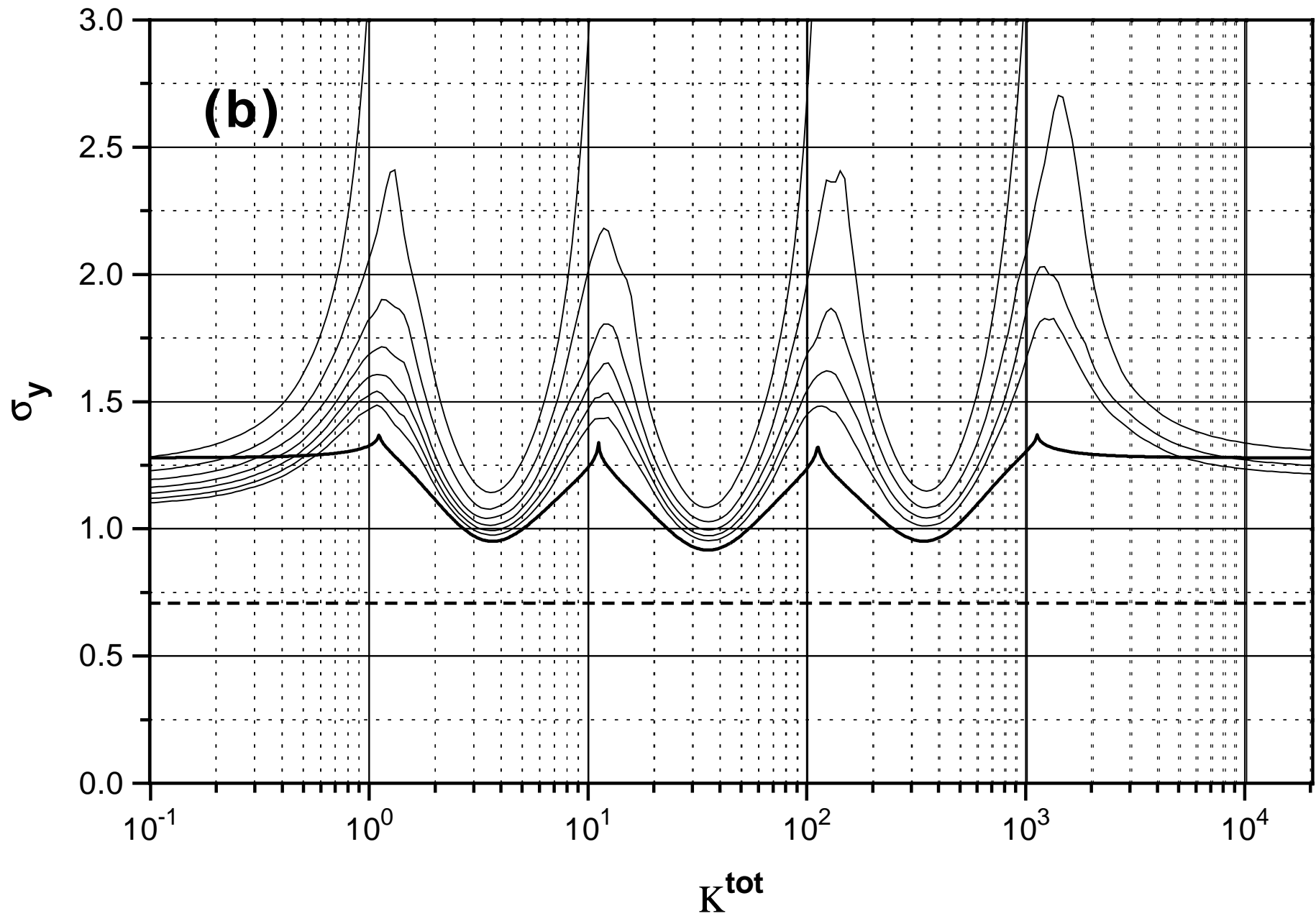
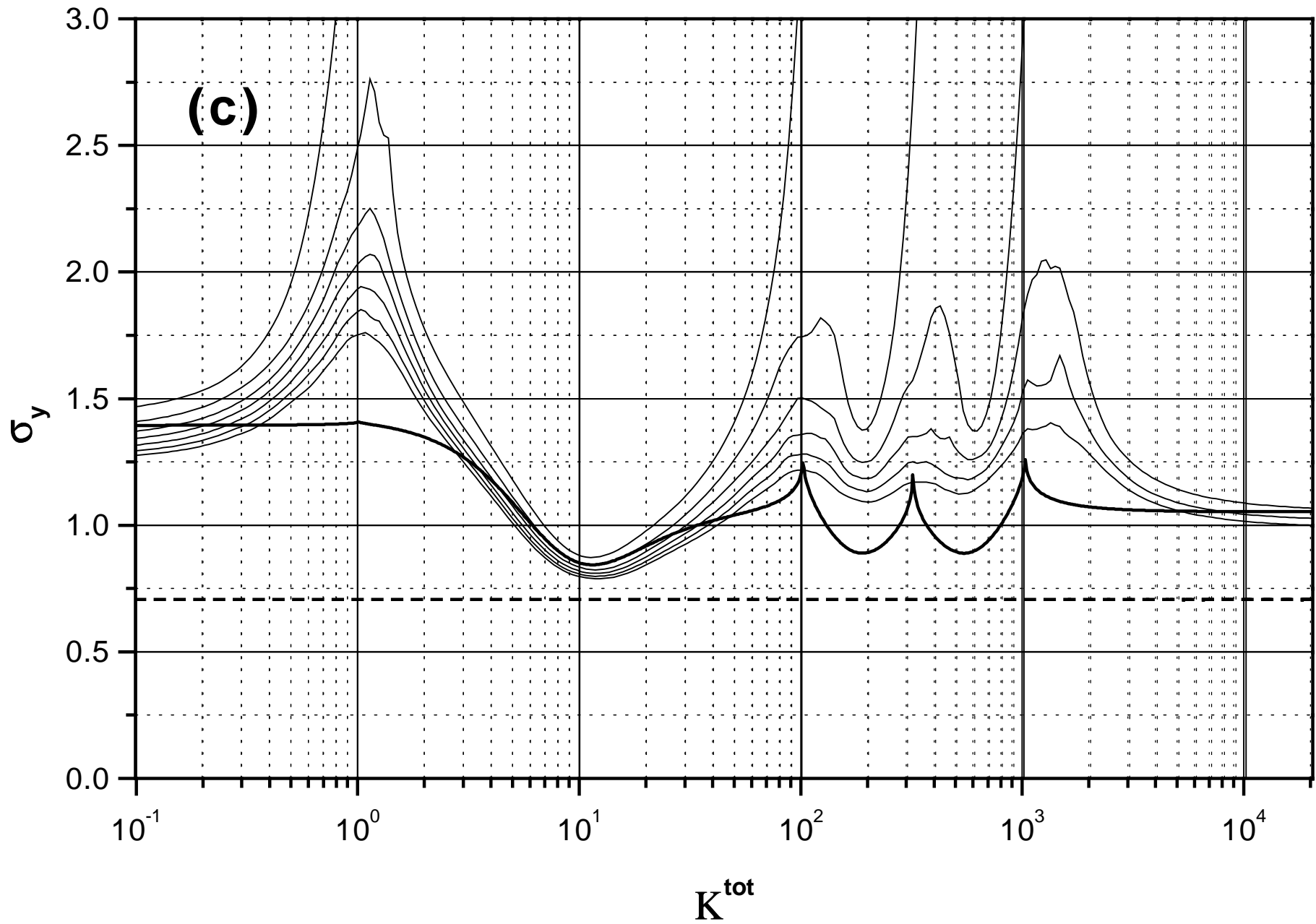
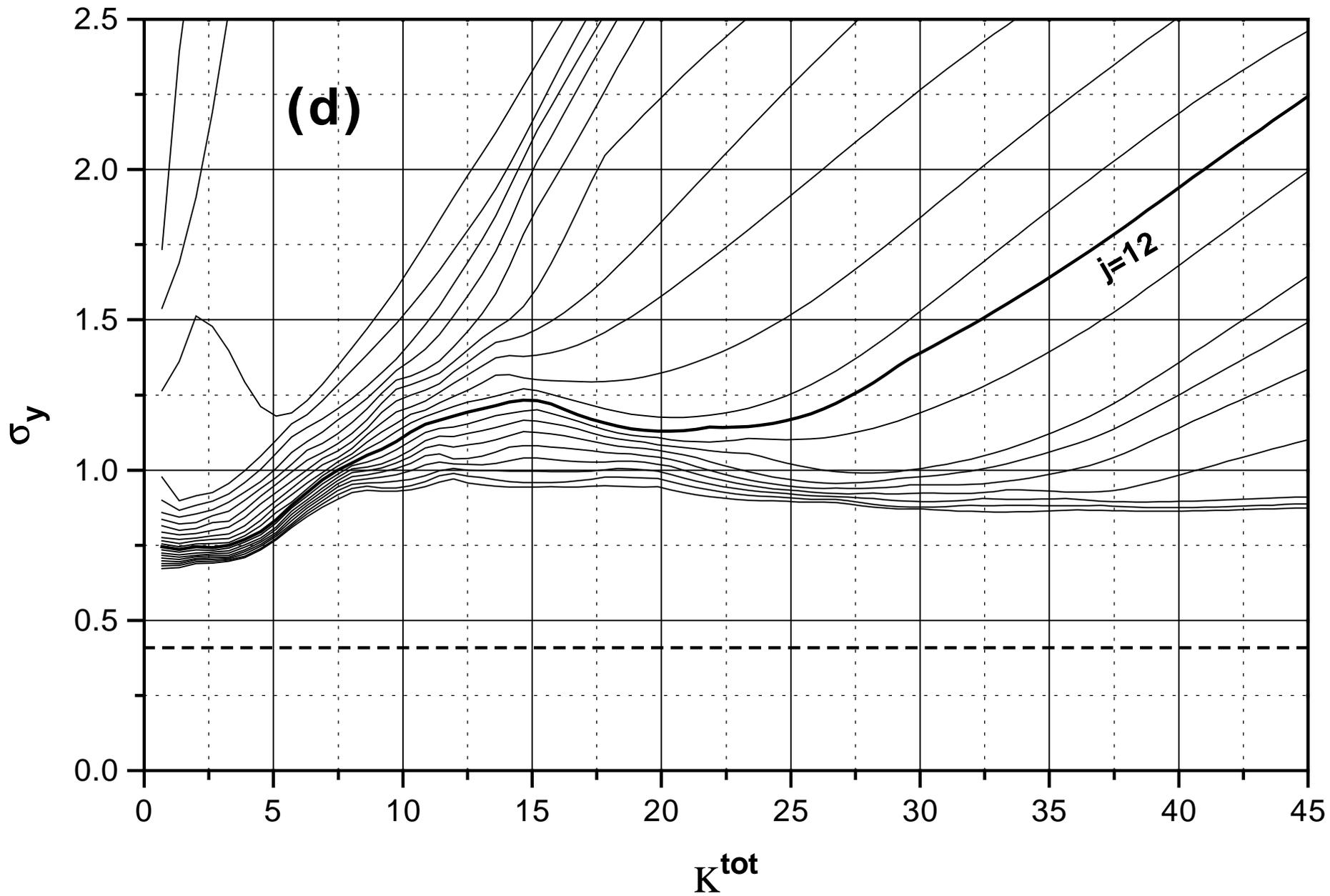
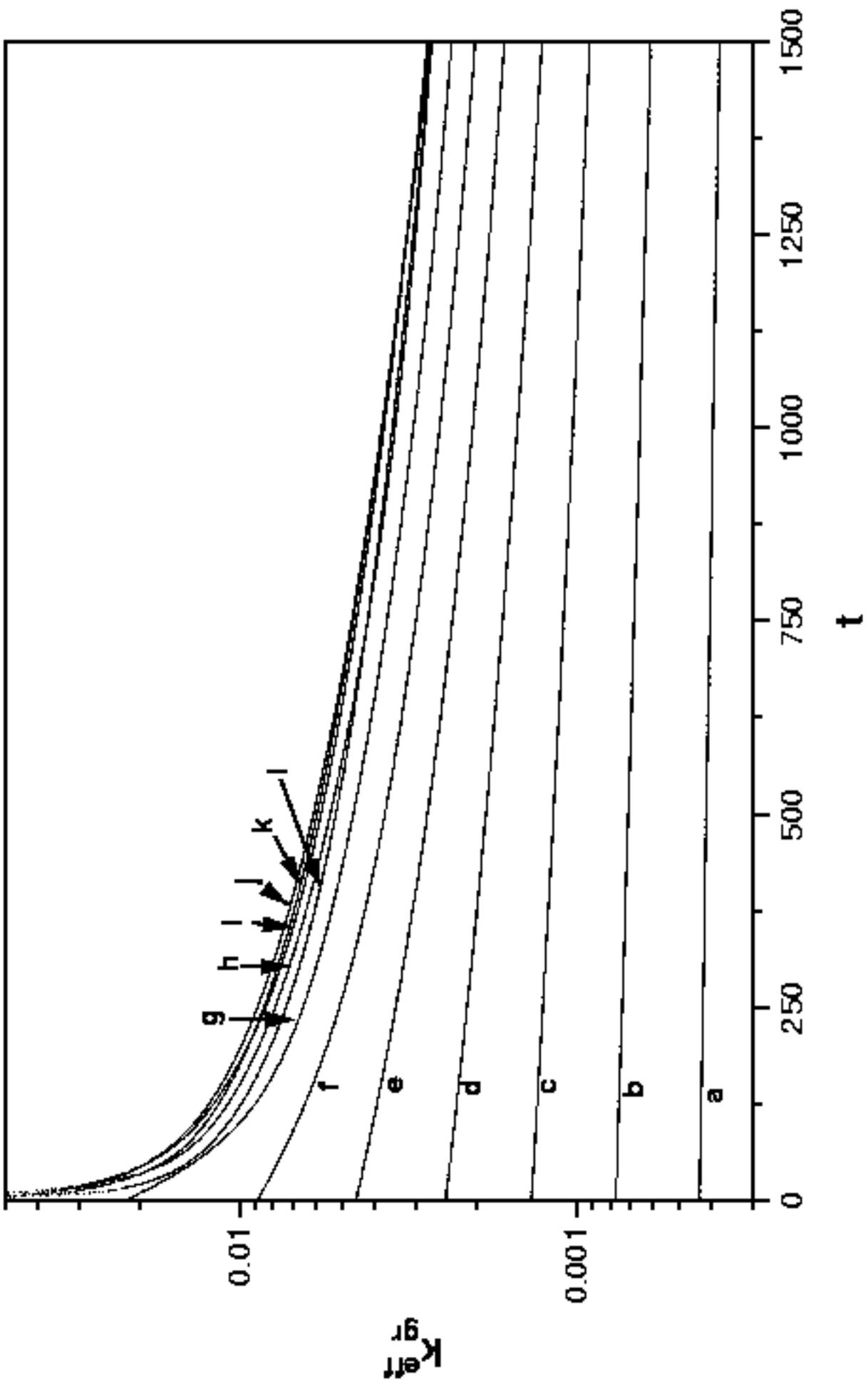
\end{document}